\documentclass[graybox]{svmult}

\usepackage{type1cm}
\usepackage{graphicx}
\usepackage{newtxtext}
\usepackage[varvw]{newtxmath}

\graphicspath{{Figures/dinnertime/}{Figures/DitherMaps/}}
\usepackage{adjustbox}
\usepackage{url}

\usepackage{tikz}
\tikzset{num/.style={anchor = center, inner sep = 2, font = \small}}
\usepackage{xcolor}
\definecolor{Emerald}{rgb}{0.31, 0.78, 0.47}
\definecolor{MidnightBlue}{rgb}{0.1, 0.1, 0.44}
\definecolor{YellowOrange}{rgb}{0.95, 0.52, 0.0}
\definecolor{Magenta}{rgb}{1.0, 0.0, 1.0}
\definecolor{Salmon}{rgb}{1.0, 0.57, 0.64}
\definecolor{LimeGreen}{rgb}{0.462745098, 0.725490196, 0.000000000}

\begin{document}
\title*{Rendering along the Hilbert Curve}
\author{Alexander Keller, Carsten W\"achter, and Nikolaus Binder}
\institute{Alexander Keller \email{akeller@nvidia.com}
\and Carsten W\"achter \email{cwachter@nvidia.com}
\and Nikolaus Binder \email{nbinder@nvidia.com} \at NVIDIA, Fasanenstr. 81, 10623 Berlin, Germany}

\maketitle

\abstract*{Based on the seminal work on Array-RQMC methods and rank-1
lattice sequences by Pierre L'Ecuyer
and collaborators, we introduce efficient deterministic algorithms for image
synthesis. Enumerating a low discrepancy sequence along the
Hilbert curve superimposed on the raster of pixels of an image, we achieve noise
characteristics that are desirable with respect to the human visual system, especially
at very low sampling rates. As compared to the state of the art,
our simple algorithms neither require randomization, nor costly
optimization, nor lookup tables. We analyze correlations of space-filling curves
and low discrepancy sequences, and demonstrate the benefits of the new algorithms in a
professional, massively parallel light transport simulation and rendering system.}

\abstract{Based on the seminal work on Array-RQMC methods and rank-1
lattice sequences by Pierre L'Ecuyer
and collaborators, we introduce efficient deterministic algorithms for image
synthesis. Enumerating a low discrepancy sequence along the
Hilbert curve superimposed on the raster of pixels of an image, we achieve noise
characteristics that are desirable with respect to the human visual system, especially
at very low sampling rates. As compared to the state of the art,
our simple algorithms neither require randomization, nor costly
optimization, nor lookup tables. We analyze correlations of space-filling curves
and low discrepancy sequences, and demonstrate the benefits of the new algorithms in a
professional, massively parallel light transport simulation and rendering system.}


\section{Introduction}

In photorealistic image synthesis by light transport simulation,
the colors of each pixel are an
integral of a high-dimensional function. While the functions
to integrate are square-integrable and hence of finite energy,
they contain discontinuities that cannot be predicted efficiently.
In practice, the pixel colors are estimated by Monte Carlo and
quasi-Monte Carlo methods sampling light transport paths that
connect light sources and cameras and summing up the contributions.

As a consequence of sampling, images appear noisy when the number of samples is insufficient.
This is quite common, when images need to be synthesized
rapidly for real-time applications and when convergence is slow
due to the intricacies of the functions to integrate. Depending on the
characteristics of the noise in the image, filtering may efficiently
improve image quality.

The success of image compression algorithms and compressive
sensing methods clearly indicates that the pixels of an image are
not independent integrals. One way to account for the correlation
of pixels is to consider image synthesis an integro-approximation
problem \cite{SampleEnum}.

In this article we propose a new way of
synthesizing images as a sequence of correlated integrals such
that noise is less perceivable by the human visual system.
We therefore review the state of the art in addressing perceived image error
in computer graphics in Sect.~\ref{Sec:HVS} and introduce our new deterministic algorithm for
image synthesis by enumerating a low discrepancy sequence along the
Hilbert Curve in Sect.~\ref{Sec:HilbertEnumeration}. We then explore extensions
for progressive image synthesis in Sect.~\ref{Sec:Progressive}
and discuss the results in Sect.~\ref{Sec:Discussion} before drawing the conclusions.

For the scope of our article, it is sufficient to understand that the mapping of
a vector of the low discrepancy sequence to a light transport path is the
same across all discussed methods and that methods only differ in which
vector of a low discrepancy sequence is assigned to what pixel. This abstraction
allows for reproducing the results. For the experiments, we use
the Iray light transport simulation and rendering system \cite{Iray_report}.
For the details we refer to
\cite{NutshellQMC,Iray_report,HashedPSF,FilterImportanceSampling}
and extensive background information in \cite{PBRT}. A recent survey
of sampling methods in computer graphics is
\cite{MyFavoriteSamples}, while the latest research focuses on
low discrepancy sequences with good low-dimensional projections
\cite{Cascaded,MatBuilder}.

\section{Visual Error in Image Synthesis} \label{Sec:HVS}

The human visual system is quite capable of recovering information
from noisy images and computer graphics has been taking advantage
of that since its early days \cite{stochcg}. Using the same set of samples
across the pixels to synthesize an image may result in disturbingly visible
aliasing artifacts. Hence, inspired by the arrangement
of receptors in a monkey's retina \cite{Yel:83}, a huge body of work
around sampling patterns with blue noise characteristics emerged.
Especially at low sampling rates and in low dimensions,
these patterns have been attractive since they are close to the ideal
of reconstructing precisely as long as the assumptions of the
sampling theorem are fulfilled, while aliases are mapped to noise,
which is very amenable to the human visual system.

For long, an important detail had not been considered explicitly: blue noise
characteristics of samples do not matter much for a single pixel integral
but when observing an ensemble of neighboring pixels. Only recently, it was
found that optimizing the parameters of a Cranley-Patterson rotation
per pixel applied to one generic set of samples can dramatically improve
the perceived image quality although the $\ell^2$-error remains about the same
\cite{DitheredSampling}. Subsequent work extends the optimization to
scramblings of the Sobol' sequence \cite{ScreenSpaceBlueNoise}.
The visual improvements have been attributed to blue noise characteristics.

Since the $\ell^2$-error between a reference image and different
sampling schemes remains about the same, the improvement in
perceived image quality must be in the the distribution of the
error and how the samples are correlated across the pixels \cite{Z-Sampler}.

Based on the family of Array-RQMC methods \cite{SortingArrayQMC} introduced by Pierre L'Ecuyer,
we propose simple deterministic quasi-Monte Carlo algorithms that result in
similar visual improvements without optimization. In addition,
an explanation for the improvements beyond blue noise characteristics is offered.
Our approach benefits from the improved uniformity of low discrepancy sequences observed when simulating
Markov chains \cite{Markov:06,SortingArrayQMC}  by ordering their states by proximity in world space. Instead
of sorting, we explore orders provided by space filling curves in screen space.

\begin{figure}
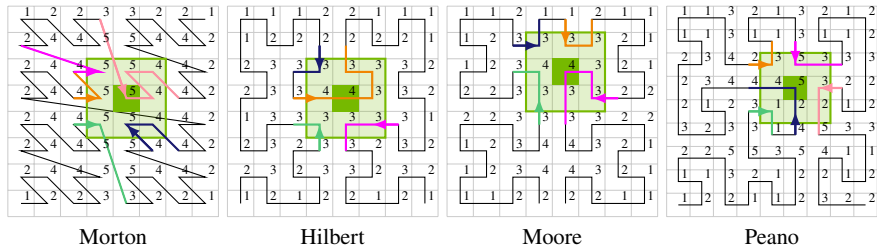

  \centering
  \begin{tabular}{cccc}
		\begin{tikzpicture}[scale = 2.8]
			\input{Figures/Curves/sfc_minicg_0.tex}
		\end{tikzpicture}&
		\begin{tikzpicture}[scale = 2.8]
			\input{Figures/Curves/sfc_minicg_1.tex}
		\end{tikzpicture}&
		\begin{tikzpicture}[scale = 2.8]
			\input{Figures/Curves/sfc_minicg_2.tex}
		\end{tikzpicture}&
		\begin{tikzpicture}[scale = 2.8]
			\input{Figures/Curves/sfc_minicg_3.tex}
		\end{tikzpicture}\\
		Morton &
		Hilbert &
		Moore &
		Peano
	\end{tabular}

  \caption{The Morton, Hilbert, Moore, and Peano space-filling curves on a pixel grid. As the Hilbert, the Moore,
  and the Peano curve only pass through neighboring pixels, they realize shortest
  routes of visiting all pixels in the sense of the Traveling Salesman problem. The pixels highlighted in dark green
  exemplify the number of segments (colored) of each space-filling curve entering the $3 \times 3$ neighborhood. This number
  of segments is depicted for each pixel and its maximum is smallest for the Hilbert and Moore curves.
  Enumerating the a low discrepancy sequence along a space filling curve, a smaller number of segments
  implies longer contiguous segments of the low discrepancy sequence used in the $3 \times 3$ neighborhood, which
  improves uniformity.
  }
  \label{Fig:SpaceFillingCurves}
\end{figure}

\section{Enumerating Pixels along the Hilbert Curve} \label{Sec:HilbertEnumeration}

Given the resolution of an image to synthesize, a deterministic low discrepancy
sequence, and a number of samples to be drawn per pixel,
our deterministic algorithm to enumerate the samples per pixel starts by
selecting the resolution of the Hilbert curve (see Fig.~\ref{Fig:SpaceFillingCurves})
to match the image resolution. We therefore determine the smallest power
of two that is larger or equal to the maximum of the image resolution in
horizontal and vertical direction. Enumerating the pixels along the
Hilbert curve, for each pixel we draw the selected samples
from the low discrepancy sequence in contiguous blocks.
Pixels outside the image are simply skipped.

As shown in Fig.~\ref{Fig:FiniteSamples}, this simple algorithm performs astonishingly
well in comparison to using the first two dimensions of the same low discrepancy sequence
to sample the pixels of the image mapped to the unit square \cite{SampleEnum}.
While both approaches expose about the same $\ell^2$-error as argued in \cite{DitheredSampling},
sampling along the Hilbert curve results in noise that is much more uniformly
distributed across the image, especially visible at low sampling rates.

The human visual system always tries to detect scale invariant features and unfortunately
finds such in the non-uniformities of noise, too. As such features are not related to the actual image
content, they are perceived as disturbing artifacts. If, however, the noise is uniform it is less
likely misinterpreted and consequently noise is not perceived as much.
As a result, one may argue that the eye is filtering the noise by integrating over areas of
uniform noise in the image. While the blue noise sampling approaches mentioned in the previous section
rely on this phenomenon, our new approaches are deterministic and
do not require optimization.

Enumerating low discrepancy sequences along space filling curves by spatial
proximity \cite{Markov:06} suggests that contiguous blocks of samples from
a low discrepancy sequence are spatially close and hence improve local uniformity.
Similarly, using a variant of the Morton curve (see Fig.~\ref{Fig:SpaceFillingCurves}) combined with scrambling
to enumerate pixels \cite{Z-Sampler} results in an error more uniformly distributed
across the image. The observable improvements are supported by
the fact that the low discrepancy of a point sequence is preserved when enumerated along
the Hilbert curve \cite{SQMC}. In addition, we can adopt an argument from \cite[Sec.3.2]{SortingArrayQMC}:
Considering an image as a line of pixels as enumerated along the Hilbert curve
and assuming the function to be integrated along the pixel to have a
gradient bounded by $K$, the total variation is bounded by $K$ times
the length of the Hilbert curve, which in turn bounds the integration error
by the Koksma-Hlawka inequality \cite{Nie:92}. While the Hilbert, Moore,
or Peano curve achieve a shortest route to connect all pixels, the Morton curve
fails to do so, which explains parts of its inferiority. On a historical note, both
\cite{Markov:06} and \cite{Z-Sampler} mention the Hilbert curve but used the Morton curve
and hence cannot not take advantage of the above argument. As
compared to the Morton and Peano curve, both the Hilbert and Moore curve expose
a smaller maximum number of curves segments in the $3 \times 3$ neighborhood
of a pixel (see Fig.~\ref{Fig:SpaceFillingCurves}), resulting in more
consecutive samples of the low discrepancy sequence
in the neighborhood, which improves uniformity locally. 

In computer graphics gradients may be bounded in parts of the integration domain,
however, such parts usually cannot be identified efficiently. Yet, Fig.~\ref{Fig:FiniteSamples}
clearly shows that in such smoother regions of an image the noise is
much more uniformly distributed when using the proposed algorithm.
The human visual system takes advantage of these local improvements.

\begin{figure}
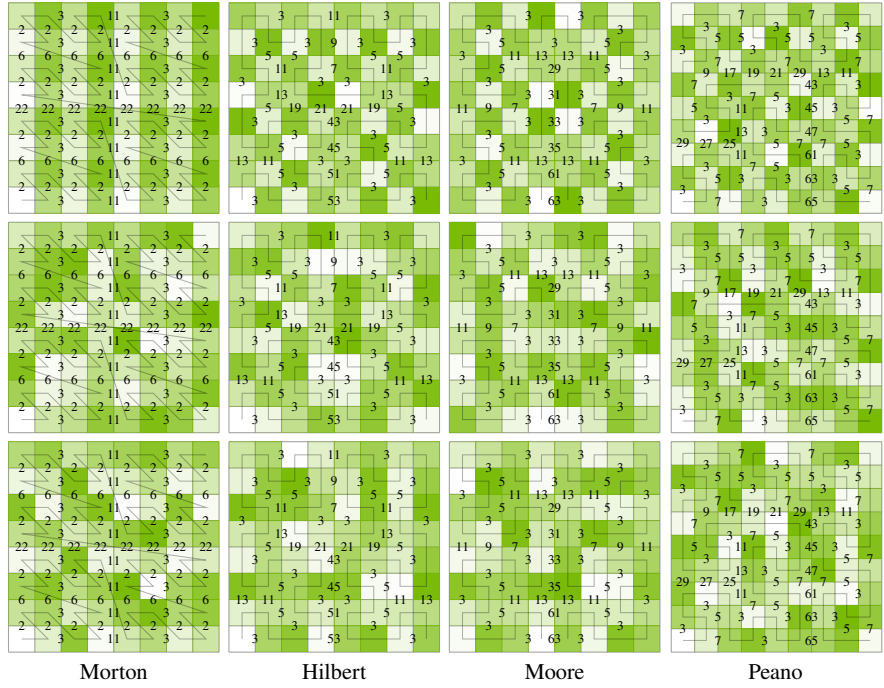

  \centering
  \begin{tabular}{cccc}
		\begin{tikzpicture}[scale = 2.8]
			\draw[opacity = 0.2, scale=1/8, shift={(-1/2, -1/2)}] (0, 0) grid (8, 8);
\draw[opacity = 0.2, scale=1/8, shift={(-1/2, -1/2)}] (0, 0) grid (8, 8);
\fill[LimeGreen, opacity=0] (-0.0625, -0.0625) rectangle ++(0.125, 0.125);
\fill[LimeGreen, opacity=0.5] (0.0625, -0.0625) rectangle ++(0.125, 0.125);
\fill[LimeGreen, opacity=0.25] (-0.0625, 0.0625) rectangle ++(0.125, 0.125);
\fill[LimeGreen, opacity=0.75] (0.0625, 0.0625) rectangle ++(0.125, 0.125);
\fill[LimeGreen, opacity=0.125] (0.1875, -0.0625) rectangle ++(0.125, 0.125);
\fill[LimeGreen, opacity=0.625] (0.3125, -0.0625) rectangle ++(0.125, 0.125);
\fill[LimeGreen, opacity=0.375] (0.1875, 0.0625) rectangle ++(0.125, 0.125);
\fill[LimeGreen, opacity=0.875] (0.3125, 0.0625) rectangle ++(0.125, 0.125);
\fill[LimeGreen, opacity=0.0625] (-0.0625, 0.1875) rectangle ++(0.125, 0.125);
\fill[LimeGreen, opacity=0.5625] (0.0625, 0.1875) rectangle ++(0.125, 0.125);
\fill[LimeGreen, opacity=0.3125] (-0.0625, 0.3125) rectangle ++(0.125, 0.125);
\fill[LimeGreen, opacity=0.8125] (0.0625, 0.3125) rectangle ++(0.125, 0.125);
\fill[LimeGreen, opacity=0.1875] (0.1875, 0.1875) rectangle ++(0.125, 0.125);
\fill[LimeGreen, opacity=0.6875] (0.3125, 0.1875) rectangle ++(0.125, 0.125);
\fill[LimeGreen, opacity=0.4375] (0.1875, 0.3125) rectangle ++(0.125, 0.125);
\fill[LimeGreen, opacity=0.9375] (0.3125, 0.3125) rectangle ++(0.125, 0.125);
\fill[LimeGreen, opacity=0.03125] (0.4375, -0.0625) rectangle ++(0.125, 0.125);
\fill[LimeGreen, opacity=0.53125] (0.5625, -0.0625) rectangle ++(0.125, 0.125);
\fill[LimeGreen, opacity=0.28125] (0.4375, 0.0625) rectangle ++(0.125, 0.125);
\fill[LimeGreen, opacity=0.78125] (0.5625, 0.0625) rectangle ++(0.125, 0.125);
\fill[LimeGreen, opacity=0.15625] (0.6875, -0.0625) rectangle ++(0.125, 0.125);
\fill[LimeGreen, opacity=0.65625] (0.8125, -0.0625) rectangle ++(0.125, 0.125);
\fill[LimeGreen, opacity=0.40625] (0.6875, 0.0625) rectangle ++(0.125, 0.125);
\fill[LimeGreen, opacity=0.90625] (0.8125, 0.0625) rectangle ++(0.125, 0.125);
\fill[LimeGreen, opacity=0.09375] (0.4375, 0.1875) rectangle ++(0.125, 0.125);
\fill[LimeGreen, opacity=0.59375] (0.5625, 0.1875) rectangle ++(0.125, 0.125);
\fill[LimeGreen, opacity=0.34375] (0.4375, 0.3125) rectangle ++(0.125, 0.125);
\fill[LimeGreen, opacity=0.84375] (0.5625, 0.3125) rectangle ++(0.125, 0.125);
\fill[LimeGreen, opacity=0.21875] (0.6875, 0.1875) rectangle ++(0.125, 0.125);
\fill[LimeGreen, opacity=0.71875] (0.8125, 0.1875) rectangle ++(0.125, 0.125);
\fill[LimeGreen, opacity=0.46875] (0.6875, 0.3125) rectangle ++(0.125, 0.125);
\fill[LimeGreen, opacity=0.96875] (0.8125, 0.3125) rectangle ++(0.125, 0.125);
\fill[LimeGreen, opacity=0.015625] (-0.0625, 0.4375) rectangle ++(0.125, 0.125);
\fill[LimeGreen, opacity=0.515625] (0.0625, 0.4375) rectangle ++(0.125, 0.125);
\fill[LimeGreen, opacity=0.265625] (-0.0625, 0.5625) rectangle ++(0.125, 0.125);
\fill[LimeGreen, opacity=0.765625] (0.0625, 0.5625) rectangle ++(0.125, 0.125);
\fill[LimeGreen, opacity=0.140625] (0.1875, 0.4375) rectangle ++(0.125, 0.125);
\fill[LimeGreen, opacity=0.640625] (0.3125, 0.4375) rectangle ++(0.125, 0.125);
\fill[LimeGreen, opacity=0.390625] (0.1875, 0.5625) rectangle ++(0.125, 0.125);
\fill[LimeGreen, opacity=0.890625] (0.3125, 0.5625) rectangle ++(0.125, 0.125);
\fill[LimeGreen, opacity=0.078125] (-0.0625, 0.6875) rectangle ++(0.125, 0.125);
\fill[LimeGreen, opacity=0.578125] (0.0625, 0.6875) rectangle ++(0.125, 0.125);
\fill[LimeGreen, opacity=0.328125] (-0.0625, 0.8125) rectangle ++(0.125, 0.125);
\fill[LimeGreen, opacity=0.828125] (0.0625, 0.8125) rectangle ++(0.125, 0.125);
\fill[LimeGreen, opacity=0.203125] (0.1875, 0.6875) rectangle ++(0.125, 0.125);
\fill[LimeGreen, opacity=0.703125] (0.3125, 0.6875) rectangle ++(0.125, 0.125);
\fill[LimeGreen, opacity=0.453125] (0.1875, 0.8125) rectangle ++(0.125, 0.125);
\fill[LimeGreen, opacity=0.953125] (0.3125, 0.8125) rectangle ++(0.125, 0.125);
\fill[LimeGreen, opacity=0.046875] (0.4375, 0.4375) rectangle ++(0.125, 0.125);
\fill[LimeGreen, opacity=0.546875] (0.5625, 0.4375) rectangle ++(0.125, 0.125);
\fill[LimeGreen, opacity=0.296875] (0.4375, 0.5625) rectangle ++(0.125, 0.125);
\fill[LimeGreen, opacity=0.796875] (0.5625, 0.5625) rectangle ++(0.125, 0.125);
\fill[LimeGreen, opacity=0.171875] (0.6875, 0.4375) rectangle ++(0.125, 0.125);
\fill[LimeGreen, opacity=0.671875] (0.8125, 0.4375) rectangle ++(0.125, 0.125);
\fill[LimeGreen, opacity=0.421875] (0.6875, 0.5625) rectangle ++(0.125, 0.125);
\fill[LimeGreen, opacity=0.921875] (0.8125, 0.5625) rectangle ++(0.125, 0.125);
\fill[LimeGreen, opacity=0.109375] (0.4375, 0.6875) rectangle ++(0.125, 0.125);
\fill[LimeGreen, opacity=0.609375] (0.5625, 0.6875) rectangle ++(0.125, 0.125);
\fill[LimeGreen, opacity=0.359375] (0.4375, 0.8125) rectangle ++(0.125, 0.125);
\fill[LimeGreen, opacity=0.859375] (0.5625, 0.8125) rectangle ++(0.125, 0.125);
\fill[LimeGreen, opacity=0.234375] (0.6875, 0.6875) rectangle ++(0.125, 0.125);
\fill[LimeGreen, opacity=0.734375] (0.8125, 0.6875) rectangle ++(0.125, 0.125);
\fill[LimeGreen, opacity=0.484375] (0.6875, 0.8125) rectangle ++(0.125, 0.125);
\fill[LimeGreen, opacity=0.984375] (0.8125, 0.8125) rectangle ++(0.125, 0.125);
\draw[opacity=0.3] (0.0, 0.0)--(0.125, 0.0)--(0.0, 0.125)--(0.125, 0.125)--(0.25, 0.0)--(0.375, 0.0)--(0.25, 0.125)--(0.375, 0.125)--(0.0, 0.25)--(0.125, 0.25)--(0.0, 0.375)--(0.125, 0.375)--(0.25, 0.25)--(0.375, 0.25)--(0.25, 0.375)--(0.375, 0.375)--(0.5, 0.0)--(0.625, 0.0)--(0.5, 0.125)--(0.625, 0.125)--(0.75, 0.0)--(0.875, 0.0)--(0.75, 0.125)--(0.875, 0.125)--(0.5, 0.25)--(0.625, 0.25)--(0.5, 0.375)--(0.625, 0.375)--(0.75, 0.25)--(0.875, 0.25)--(0.75, 0.375)--(0.875, 0.375)--(0.0, 0.5)--(0.125, 0.5)--(0.0, 0.625)--(0.125, 0.625)--(0.25, 0.5)--(0.375, 0.5)--(0.25, 0.625)--(0.375, 0.625)--(0.0, 0.75)--(0.125, 0.75)--(0.0, 0.875)--(0.125, 0.875)--(0.25, 0.75)--(0.375, 0.75)--(0.25, 0.875)--(0.375, 0.875)--(0.5, 0.5)--(0.625, 0.5)--(0.5, 0.625)--(0.625, 0.625)--(0.75, 0.5)--(0.875, 0.5)--(0.75, 0.625)--(0.875, 0.625)--(0.5, 0.75)--(0.625, 0.75)--(0.5, 0.875)--(0.625, 0.875)--(0.75, 0.75)--(0.875, 0.75)--(0.75, 0.875)--(0.875, 0.875);
\node[font=\tiny] at (0.0, 0.0625) { 2 };
\node[font=\tiny] at (0.0, 0.1875) { 6 };
\node[font=\tiny] at (0.0, 0.3125) { 2 };
\node[font=\tiny] at (0.0, 0.4375) { 22 };
\node[font=\tiny] at (0.0, 0.5625) { 2 };
\node[font=\tiny] at (0.0, 0.6875) { 6 };
\node[font=\tiny] at (0.0, 0.8125) { 2 };
\node[font=\tiny] at (0.125, 0.0625) { 2 };
\node[font=\tiny] at (0.125, 0.1875) { 6 };
\node[font=\tiny] at (0.125, 0.3125) { 2 };
\node[font=\tiny] at (0.125, 0.4375) { 22 };
\node[font=\tiny] at (0.125, 0.5625) { 2 };
\node[font=\tiny] at (0.125, 0.6875) { 6 };
\node[font=\tiny] at (0.125, 0.8125) { 2 };
\node[font=\tiny] at (0.1875, 0.0) { 3 };
\node[font=\tiny] at (0.1875, 0.125) { 3 };
\node[font=\tiny] at (0.25, 0.0625) { 2 };
\node[font=\tiny] at (0.1875, 0.25) { 3 };
\node[font=\tiny] at (0.25, 0.1875) { 6 };
\node[font=\tiny] at (0.1875, 0.375) { 3 };
\node[font=\tiny] at (0.25, 0.3125) { 2 };
\node[font=\tiny] at (0.1875, 0.5) { 3 };
\node[font=\tiny] at (0.25, 0.4375) { 22 };
\node[font=\tiny] at (0.1875, 0.625) { 3 };
\node[font=\tiny] at (0.25, 0.5625) { 2 };
\node[font=\tiny] at (0.1875, 0.75) { 3 };
\node[font=\tiny] at (0.25, 0.6875) { 6 };
\node[font=\tiny] at (0.1875, 0.875) { 3 };
\node[font=\tiny] at (0.25, 0.8125) { 2 };
\node[font=\tiny] at (0.375, 0.0625) { 2 };
\node[font=\tiny] at (0.375, 0.1875) { 6 };
\node[font=\tiny] at (0.375, 0.3125) { 2 };
\node[font=\tiny] at (0.375, 0.4375) { 22 };
\node[font=\tiny] at (0.375, 0.5625) { 2 };
\node[font=\tiny] at (0.375, 0.6875) { 6 };
\node[font=\tiny] at (0.375, 0.8125) { 2 };
\node[font=\tiny] at (0.4375, 0.0) { 11 };
\node[font=\tiny] at (0.4375, 0.125) { 11 };
\node[font=\tiny] at (0.5, 0.0625) { 2 };
\node[font=\tiny] at (0.4375, 0.25) { 11 };
\node[font=\tiny] at (0.5, 0.1875) { 6 };
\node[font=\tiny] at (0.4375, 0.375) { 11 };
\node[font=\tiny] at (0.5, 0.3125) { 2 };
\node[font=\tiny] at (0.4375, 0.5) { 11 };
\node[font=\tiny] at (0.5, 0.4375) { 22 };
\node[font=\tiny] at (0.4375, 0.625) { 11 };
\node[font=\tiny] at (0.5, 0.5625) { 2 };
\node[font=\tiny] at (0.4375, 0.75) { 11 };
\node[font=\tiny] at (0.5, 0.6875) { 6 };
\node[font=\tiny] at (0.4375, 0.875) { 11 };
\node[font=\tiny] at (0.5, 0.8125) { 2 };
\node[font=\tiny] at (0.625, 0.0625) { 2 };
\node[font=\tiny] at (0.625, 0.1875) { 6 };
\node[font=\tiny] at (0.625, 0.3125) { 2 };
\node[font=\tiny] at (0.625, 0.4375) { 22 };
\node[font=\tiny] at (0.625, 0.5625) { 2 };
\node[font=\tiny] at (0.625, 0.6875) { 6 };
\node[font=\tiny] at (0.625, 0.8125) { 2 };
\node[font=\tiny] at (0.6875, 0.0) { 3 };
\node[font=\tiny] at (0.6875, 0.125) { 3 };
\node[font=\tiny] at (0.75, 0.0625) { 2 };
\node[font=\tiny] at (0.6875, 0.25) { 3 };
\node[font=\tiny] at (0.75, 0.1875) { 6 };
\node[font=\tiny] at (0.6875, 0.375) { 3 };
\node[font=\tiny] at (0.75, 0.3125) { 2 };
\node[font=\tiny] at (0.6875, 0.5) { 3 };
\node[font=\tiny] at (0.75, 0.4375) { 22 };
\node[font=\tiny] at (0.6875, 0.625) { 3 };
\node[font=\tiny] at (0.75, 0.5625) { 2 };
\node[font=\tiny] at (0.6875, 0.75) { 3 };
\node[font=\tiny] at (0.75, 0.6875) { 6 };
\node[font=\tiny] at (0.6875, 0.875) { 3 };
\node[font=\tiny] at (0.75, 0.8125) { 2 };
\node[font=\tiny] at (0.875, 0.0625) { 2 };
\node[font=\tiny] at (0.875, 0.1875) { 6 };
\node[font=\tiny] at (0.875, 0.3125) { 2 };
\node[font=\tiny] at (0.875, 0.4375) { 22 };
\node[font=\tiny] at (0.875, 0.5625) { 2 };
\node[font=\tiny] at (0.875, 0.6875) { 6 };
\node[font=\tiny] at (0.875, 0.8125) { 2 };
		\end{tikzpicture}
		& \begin{tikzpicture}[scale = 2.8]
			\draw[opacity = 0.2, scale=1/8, shift={(-1/2, -1/2)}] (0, 0) grid (8, 8);
\draw[opacity = 0.2, scale=1/8, shift={(-1/2, -1/2)}] (0, 0) grid (8, 8);
\fill[LimeGreen, opacity=0] (-0.0625, -0.0625) rectangle ++(0.125, 0.125);
\fill[LimeGreen, opacity=0.5] (-0.0625, 0.0625) rectangle ++(0.125, 0.125);
\fill[LimeGreen, opacity=0.25] (0.0625, 0.0625) rectangle ++(0.125, 0.125);
\fill[LimeGreen, opacity=0.75] (0.0625, -0.0625) rectangle ++(0.125, 0.125);
\fill[LimeGreen, opacity=0.125] (0.1875, -0.0625) rectangle ++(0.125, 0.125);
\fill[LimeGreen, opacity=0.625] (0.3125, -0.0625) rectangle ++(0.125, 0.125);
\fill[LimeGreen, opacity=0.375] (0.3125, 0.0625) rectangle ++(0.125, 0.125);
\fill[LimeGreen, opacity=0.875] (0.1875, 0.0625) rectangle ++(0.125, 0.125);
\fill[LimeGreen, opacity=0.0625] (0.1875, 0.1875) rectangle ++(0.125, 0.125);
\fill[LimeGreen, opacity=0.5625] (0.3125, 0.1875) rectangle ++(0.125, 0.125);
\fill[LimeGreen, opacity=0.3125] (0.3125, 0.3125) rectangle ++(0.125, 0.125);
\fill[LimeGreen, opacity=0.8125] (0.1875, 0.3125) rectangle ++(0.125, 0.125);
\fill[LimeGreen, opacity=0.1875] (0.0625, 0.3125) rectangle ++(0.125, 0.125);
\fill[LimeGreen, opacity=0.6875] (0.0625, 0.1875) rectangle ++(0.125, 0.125);
\fill[LimeGreen, opacity=0.4375] (-0.0625, 0.1875) rectangle ++(0.125, 0.125);
\fill[LimeGreen, opacity=0.9375] (-0.0625, 0.3125) rectangle ++(0.125, 0.125);
\fill[LimeGreen, opacity=0.03125] (-0.0625, 0.4375) rectangle ++(0.125, 0.125);
\fill[LimeGreen, opacity=0.53125] (0.0625, 0.4375) rectangle ++(0.125, 0.125);
\fill[LimeGreen, opacity=0.28125] (0.0625, 0.5625) rectangle ++(0.125, 0.125);
\fill[LimeGreen, opacity=0.78125] (-0.0625, 0.5625) rectangle ++(0.125, 0.125);
\fill[LimeGreen, opacity=0.15625] (-0.0625, 0.6875) rectangle ++(0.125, 0.125);
\fill[LimeGreen, opacity=0.65625] (-0.0625, 0.8125) rectangle ++(0.125, 0.125);
\fill[LimeGreen, opacity=0.40625] (0.0625, 0.8125) rectangle ++(0.125, 0.125);
\fill[LimeGreen, opacity=0.90625] (0.0625, 0.6875) rectangle ++(0.125, 0.125);
\fill[LimeGreen, opacity=0.09375] (0.1875, 0.6875) rectangle ++(0.125, 0.125);
\fill[LimeGreen, opacity=0.59375] (0.1875, 0.8125) rectangle ++(0.125, 0.125);
\fill[LimeGreen, opacity=0.34375] (0.3125, 0.8125) rectangle ++(0.125, 0.125);
\fill[LimeGreen, opacity=0.84375] (0.3125, 0.6875) rectangle ++(0.125, 0.125);
\fill[LimeGreen, opacity=0.21875] (0.3125, 0.5625) rectangle ++(0.125, 0.125);
\fill[LimeGreen, opacity=0.71875] (0.1875, 0.5625) rectangle ++(0.125, 0.125);
\fill[LimeGreen, opacity=0.46875] (0.1875, 0.4375) rectangle ++(0.125, 0.125);
\fill[LimeGreen, opacity=0.96875] (0.3125, 0.4375) rectangle ++(0.125, 0.125);
\fill[LimeGreen, opacity=0.015625] (0.4375, 0.4375) rectangle ++(0.125, 0.125);
\fill[LimeGreen, opacity=0.515625] (0.5625, 0.4375) rectangle ++(0.125, 0.125);
\fill[LimeGreen, opacity=0.265625] (0.5625, 0.5625) rectangle ++(0.125, 0.125);
\fill[LimeGreen, opacity=0.765625] (0.4375, 0.5625) rectangle ++(0.125, 0.125);
\fill[LimeGreen, opacity=0.140625] (0.4375, 0.6875) rectangle ++(0.125, 0.125);
\fill[LimeGreen, opacity=0.640625] (0.4375, 0.8125) rectangle ++(0.125, 0.125);
\fill[LimeGreen, opacity=0.390625] (0.5625, 0.8125) rectangle ++(0.125, 0.125);
\fill[LimeGreen, opacity=0.890625] (0.5625, 0.6875) rectangle ++(0.125, 0.125);
\fill[LimeGreen, opacity=0.078125] (0.6875, 0.6875) rectangle ++(0.125, 0.125);
\fill[LimeGreen, opacity=0.578125] (0.6875, 0.8125) rectangle ++(0.125, 0.125);
\fill[LimeGreen, opacity=0.328125] (0.8125, 0.8125) rectangle ++(0.125, 0.125);
\fill[LimeGreen, opacity=0.828125] (0.8125, 0.6875) rectangle ++(0.125, 0.125);
\fill[LimeGreen, opacity=0.203125] (0.8125, 0.5625) rectangle ++(0.125, 0.125);
\fill[LimeGreen, opacity=0.703125] (0.6875, 0.5625) rectangle ++(0.125, 0.125);
\fill[LimeGreen, opacity=0.453125] (0.6875, 0.4375) rectangle ++(0.125, 0.125);
\fill[LimeGreen, opacity=0.953125] (0.8125, 0.4375) rectangle ++(0.125, 0.125);
\fill[LimeGreen, opacity=0.046875] (0.8125, 0.3125) rectangle ++(0.125, 0.125);
\fill[LimeGreen, opacity=0.546875] (0.8125, 0.1875) rectangle ++(0.125, 0.125);
\fill[LimeGreen, opacity=0.296875] (0.6875, 0.1875) rectangle ++(0.125, 0.125);
\fill[LimeGreen, opacity=0.796875] (0.6875, 0.3125) rectangle ++(0.125, 0.125);
\fill[LimeGreen, opacity=0.171875] (0.5625, 0.3125) rectangle ++(0.125, 0.125);
\fill[LimeGreen, opacity=0.671875] (0.4375, 0.3125) rectangle ++(0.125, 0.125);
\fill[LimeGreen, opacity=0.421875] (0.4375, 0.1875) rectangle ++(0.125, 0.125);
\fill[LimeGreen, opacity=0.921875] (0.5625, 0.1875) rectangle ++(0.125, 0.125);
\fill[LimeGreen, opacity=0.109375] (0.5625, 0.0625) rectangle ++(0.125, 0.125);
\fill[LimeGreen, opacity=0.609375] (0.4375, 0.0625) rectangle ++(0.125, 0.125);
\fill[LimeGreen, opacity=0.359375] (0.4375, -0.0625) rectangle ++(0.125, 0.125);
\fill[LimeGreen, opacity=0.859375] (0.5625, -0.0625) rectangle ++(0.125, 0.125);
\fill[LimeGreen, opacity=0.234375] (0.6875, -0.0625) rectangle ++(0.125, 0.125);
\fill[LimeGreen, opacity=0.734375] (0.6875, 0.0625) rectangle ++(0.125, 0.125);
\fill[LimeGreen, opacity=0.484375] (0.8125, 0.0625) rectangle ++(0.125, 0.125);
\fill[LimeGreen, opacity=0.984375] (0.8125, -0.0625) rectangle ++(0.125, 0.125);
\draw[opacity=0.3] (0.0, 0.0)--(0.0, 0.125)--(0.125, 0.125)--(0.125, 0.0)--(0.25, 0.0)--(0.375, 0.0)--(0.375, 0.125)--(0.25, 0.125)--(0.25, 0.25)--(0.375, 0.25)--(0.375, 0.375)--(0.25, 0.375)--(0.125, 0.375)--(0.125, 0.25)--(0.0, 0.25)--(0.0, 0.375)--(0.0, 0.5)--(0.125, 0.5)--(0.125, 0.625)--(0.0, 0.625)--(0.0, 0.75)--(0.0, 0.875)--(0.125, 0.875)--(0.125, 0.75)--(0.25, 0.75)--(0.25, 0.875)--(0.375, 0.875)--(0.375, 0.75)--(0.375, 0.625)--(0.25, 0.625)--(0.25, 0.5)--(0.375, 0.5)--(0.5, 0.5)--(0.625, 0.5)--(0.625, 0.625)--(0.5, 0.625)--(0.5, 0.75)--(0.5, 0.875)--(0.625, 0.875)--(0.625, 0.75)--(0.75, 0.75)--(0.75, 0.875)--(0.875, 0.875)--(0.875, 0.75)--(0.875, 0.625)--(0.75, 0.625)--(0.75, 0.5)--(0.875, 0.5)--(0.875, 0.375)--(0.875, 0.25)--(0.75, 0.25)--(0.75, 0.375)--(0.625, 0.375)--(0.5, 0.375)--(0.5, 0.25)--(0.625, 0.25)--(0.625, 0.125)--(0.5, 0.125)--(0.5, 0.0)--(0.625, 0.0)--(0.75, 0.0)--(0.75, 0.125)--(0.875, 0.125)--(0.875, 0.0);
\node[font=\tiny] at (0.0, 0.1875) { 13 };
\node[font=\tiny] at (0.0, 0.5625) { 3 };
\node[font=\tiny] at (0.0625, 0.0) { 3 };
\node[font=\tiny] at (0.125, 0.1875) { 11 };
\node[font=\tiny] at (0.0625, 0.375) { 3 };
\node[font=\tiny] at (0.125, 0.4375) { 5 };
\node[font=\tiny] at (0.0625, 0.75) { 3 };
\node[font=\tiny] at (0.125, 0.6875) { 5 };
\node[font=\tiny] at (0.1875, 0.125) { 5 };
\node[font=\tiny] at (0.25, 0.0625) { 3 };
\node[font=\tiny] at (0.1875, 0.25) { 5 };
\node[font=\tiny] at (0.25, 0.3125) { 3 };
\node[font=\tiny] at (0.1875, 0.5) { 13 };
\node[font=\tiny] at (0.25, 0.4375) { 19 };
\node[font=\tiny] at (0.1875, 0.625) { 11 };
\node[font=\tiny] at (0.25, 0.6875) { 5 };
\node[font=\tiny] at (0.1875, 0.875) { 3 };
\node[font=\tiny] at (0.375, 0.1875) { 3 };
\node[font=\tiny] at (0.375, 0.4375) { 21 };
\node[font=\tiny] at (0.375, 0.5625) { 3 };
\node[font=\tiny] at (0.3125, 0.75) { 3 };
\node[font=\tiny] at (0.4375, 0.0) { 53 };
\node[font=\tiny] at (0.4375, 0.125) { 51 };
\node[font=\tiny] at (0.4375, 0.25) { 45 };
\node[font=\tiny] at (0.5, 0.1875) { 3 };
\node[font=\tiny] at (0.4375, 0.375) { 43 };
\node[font=\tiny] at (0.5, 0.4375) { 21 };
\node[font=\tiny] at (0.4375, 0.625) { 7 };
\node[font=\tiny] at (0.5, 0.5625) { 3 };
\node[font=\tiny] at (0.4375, 0.75) { 9 };
\node[font=\tiny] at (0.4375, 0.875) { 11 };
\node[font=\tiny] at (0.625, 0.0625) { 3 };
\node[font=\tiny] at (0.625, 0.3125) { 3 };
\node[font=\tiny] at (0.625, 0.4375) { 19 };
\node[font=\tiny] at (0.5625, 0.75) { 3 };
\node[font=\tiny] at (0.625, 0.6875) { 5 };
\node[font=\tiny] at (0.6875, 0.125) { 5 };
\node[font=\tiny] at (0.6875, 0.25) { 5 };
\node[font=\tiny] at (0.75, 0.1875) { 11 };
\node[font=\tiny] at (0.6875, 0.5) { 13 };
\node[font=\tiny] at (0.75, 0.4375) { 5 };
\node[font=\tiny] at (0.6875, 0.625) { 11 };
\node[font=\tiny] at (0.75, 0.6875) { 5 };
\node[font=\tiny] at (0.6875, 0.875) { 3 };
\node[font=\tiny] at (0.8125, 0.0) { 3 };
\node[font=\tiny] at (0.875, 0.1875) { 13 };
\node[font=\tiny] at (0.8125, 0.375) { 3 };
\node[font=\tiny] at (0.875, 0.5625) { 3 };
\node[font=\tiny] at (0.8125, 0.75) { 3 };
		\end{tikzpicture}
		& \begin{tikzpicture}[scale = 2.8]
			\draw[opacity = 0.2, scale=1/8, shift={(-1/2, -1/2)}] (0, 0) grid (8, 8);
\draw[opacity = 0.2, scale=1/8, shift={(-1/2, -1/2)}] (0, 0) grid (8, 8);
\fill[LimeGreen, opacity=0] (0.3125, -0.0625) rectangle ++(0.125, 0.125);
\fill[LimeGreen, opacity=0.5] (0.3125, 0.0625) rectangle ++(0.125, 0.125);
\fill[LimeGreen, opacity=0.25] (0.1875, 0.0625) rectangle ++(0.125, 0.125);
\fill[LimeGreen, opacity=0.75] (0.1875, -0.0625) rectangle ++(0.125, 0.125);
\fill[LimeGreen, opacity=0.125] (0.0625, -0.0625) rectangle ++(0.125, 0.125);
\fill[LimeGreen, opacity=0.625] (-0.0625, -0.0625) rectangle ++(0.125, 0.125);
\fill[LimeGreen, opacity=0.375] (-0.0625, 0.0625) rectangle ++(0.125, 0.125);
\fill[LimeGreen, opacity=0.875] (0.0625, 0.0625) rectangle ++(0.125, 0.125);
\fill[LimeGreen, opacity=0.0625] (0.0625, 0.1875) rectangle ++(0.125, 0.125);
\fill[LimeGreen, opacity=0.5625] (-0.0625, 0.1875) rectangle ++(0.125, 0.125);
\fill[LimeGreen, opacity=0.3125] (-0.0625, 0.3125) rectangle ++(0.125, 0.125);
\fill[LimeGreen, opacity=0.8125] (0.0625, 0.3125) rectangle ++(0.125, 0.125);
\fill[LimeGreen, opacity=0.1875] (0.1875, 0.3125) rectangle ++(0.125, 0.125);
\fill[LimeGreen, opacity=0.6875] (0.1875, 0.1875) rectangle ++(0.125, 0.125);
\fill[LimeGreen, opacity=0.4375] (0.3125, 0.1875) rectangle ++(0.125, 0.125);
\fill[LimeGreen, opacity=0.9375] (0.3125, 0.3125) rectangle ++(0.125, 0.125);
\fill[LimeGreen, opacity=0.03125] (0.3125, 0.4375) rectangle ++(0.125, 0.125);
\fill[LimeGreen, opacity=0.53125] (0.3125, 0.5625) rectangle ++(0.125, 0.125);
\fill[LimeGreen, opacity=0.28125] (0.1875, 0.5625) rectangle ++(0.125, 0.125);
\fill[LimeGreen, opacity=0.78125] (0.1875, 0.4375) rectangle ++(0.125, 0.125);
\fill[LimeGreen, opacity=0.15625] (0.0625, 0.4375) rectangle ++(0.125, 0.125);
\fill[LimeGreen, opacity=0.65625] (-0.0625, 0.4375) rectangle ++(0.125, 0.125);
\fill[LimeGreen, opacity=0.40625] (-0.0625, 0.5625) rectangle ++(0.125, 0.125);
\fill[LimeGreen, opacity=0.90625] (0.0625, 0.5625) rectangle ++(0.125, 0.125);
\fill[LimeGreen, opacity=0.09375] (0.0625, 0.6875) rectangle ++(0.125, 0.125);
\fill[LimeGreen, opacity=0.59375] (-0.0625, 0.6875) rectangle ++(0.125, 0.125);
\fill[LimeGreen, opacity=0.34375] (-0.0625, 0.8125) rectangle ++(0.125, 0.125);
\fill[LimeGreen, opacity=0.84375] (0.0625, 0.8125) rectangle ++(0.125, 0.125);
\fill[LimeGreen, opacity=0.21875] (0.1875, 0.8125) rectangle ++(0.125, 0.125);
\fill[LimeGreen, opacity=0.71875] (0.1875, 0.6875) rectangle ++(0.125, 0.125);
\fill[LimeGreen, opacity=0.46875] (0.3125, 0.6875) rectangle ++(0.125, 0.125);
\fill[LimeGreen, opacity=0.96875] (0.3125, 0.8125) rectangle ++(0.125, 0.125);
\fill[LimeGreen, opacity=0.015625] (0.4375, 0.8125) rectangle ++(0.125, 0.125);
\fill[LimeGreen, opacity=0.515625] (0.4375, 0.6875) rectangle ++(0.125, 0.125);
\fill[LimeGreen, opacity=0.265625] (0.5625, 0.6875) rectangle ++(0.125, 0.125);
\fill[LimeGreen, opacity=0.765625] (0.5625, 0.8125) rectangle ++(0.125, 0.125);
\fill[LimeGreen, opacity=0.140625] (0.6875, 0.8125) rectangle ++(0.125, 0.125);
\fill[LimeGreen, opacity=0.640625] (0.8125, 0.8125) rectangle ++(0.125, 0.125);
\fill[LimeGreen, opacity=0.390625] (0.8125, 0.6875) rectangle ++(0.125, 0.125);
\fill[LimeGreen, opacity=0.890625] (0.6875, 0.6875) rectangle ++(0.125, 0.125);
\fill[LimeGreen, opacity=0.078125] (0.6875, 0.5625) rectangle ++(0.125, 0.125);
\fill[LimeGreen, opacity=0.578125] (0.8125, 0.5625) rectangle ++(0.125, 0.125);
\fill[LimeGreen, opacity=0.328125] (0.8125, 0.4375) rectangle ++(0.125, 0.125);
\fill[LimeGreen, opacity=0.828125] (0.6875, 0.4375) rectangle ++(0.125, 0.125);
\fill[LimeGreen, opacity=0.203125] (0.5625, 0.4375) rectangle ++(0.125, 0.125);
\fill[LimeGreen, opacity=0.703125] (0.5625, 0.5625) rectangle ++(0.125, 0.125);
\fill[LimeGreen, opacity=0.453125] (0.4375, 0.5625) rectangle ++(0.125, 0.125);
\fill[LimeGreen, opacity=0.953125] (0.4375, 0.4375) rectangle ++(0.125, 0.125);
\fill[LimeGreen, opacity=0.046875] (0.4375, 0.3125) rectangle ++(0.125, 0.125);
\fill[LimeGreen, opacity=0.546875] (0.4375, 0.1875) rectangle ++(0.125, 0.125);
\fill[LimeGreen, opacity=0.296875] (0.5625, 0.1875) rectangle ++(0.125, 0.125);
\fill[LimeGreen, opacity=0.796875] (0.5625, 0.3125) rectangle ++(0.125, 0.125);
\fill[LimeGreen, opacity=0.171875] (0.6875, 0.3125) rectangle ++(0.125, 0.125);
\fill[LimeGreen, opacity=0.671875] (0.8125, 0.3125) rectangle ++(0.125, 0.125);
\fill[LimeGreen, opacity=0.421875] (0.8125, 0.1875) rectangle ++(0.125, 0.125);
\fill[LimeGreen, opacity=0.921875] (0.6875, 0.1875) rectangle ++(0.125, 0.125);
\fill[LimeGreen, opacity=0.109375] (0.6875, 0.0625) rectangle ++(0.125, 0.125);
\fill[LimeGreen, opacity=0.609375] (0.8125, 0.0625) rectangle ++(0.125, 0.125);
\fill[LimeGreen, opacity=0.359375] (0.8125, -0.0625) rectangle ++(0.125, 0.125);
\fill[LimeGreen, opacity=0.859375] (0.6875, -0.0625) rectangle ++(0.125, 0.125);
\fill[LimeGreen, opacity=0.234375] (0.5625, -0.0625) rectangle ++(0.125, 0.125);
\fill[LimeGreen, opacity=0.734375] (0.5625, 0.0625) rectangle ++(0.125, 0.125);
\fill[LimeGreen, opacity=0.484375] (0.4375, 0.0625) rectangle ++(0.125, 0.125);
\fill[LimeGreen, opacity=0.984375] (0.4375, -0.0625) rectangle ++(0.125, 0.125);
\draw[opacity=0.3] (0.375, 0.0)--(0.375, 0.125)--(0.25, 0.125)--(0.25, 0.0)--(0.125, 0.0)--(0.0, 0.0)--(0.0, 0.125)--(0.125, 0.125)--(0.125, 0.25)--(0.0, 0.25)--(0.0, 0.375)--(0.125, 0.375)--(0.25, 0.375)--(0.25, 0.25)--(0.375, 0.25)--(0.375, 0.375)--(0.375, 0.5)--(0.375, 0.625)--(0.25, 0.625)--(0.25, 0.5)--(0.125, 0.5)--(0.0, 0.5)--(0.0, 0.625)--(0.125, 0.625)--(0.125, 0.75)--(0.0, 0.75)--(0.0, 0.875)--(0.125, 0.875)--(0.25, 0.875)--(0.25, 0.75)--(0.375, 0.75)--(0.375, 0.875)--(0.5, 0.875)--(0.5, 0.75)--(0.625, 0.75)--(0.625, 0.875)--(0.75, 0.875)--(0.875, 0.875)--(0.875, 0.75)--(0.75, 0.75)--(0.75, 0.625)--(0.875, 0.625)--(0.875, 0.5)--(0.75, 0.5)--(0.625, 0.5)--(0.625, 0.625)--(0.5, 0.625)--(0.5, 0.5)--(0.5, 0.375)--(0.5, 0.25)--(0.625, 0.25)--(0.625, 0.375)--(0.75, 0.375)--(0.875, 0.375)--(0.875, 0.25)--(0.75, 0.25)--(0.75, 0.125)--(0.875, 0.125)--(0.875, 0.0)--(0.75, 0.0)--(0.625, 0.0)--(0.625, 0.125)--(0.5, 0.125)--(0.5, 0.0);
\node[font=\tiny] at (0.0, 0.1875) { 3 };
\node[font=\tiny] at (0.0, 0.4375) { 11 };
\node[font=\tiny] at (0.0, 0.6875) { 3 };
\node[font=\tiny] at (0.125, 0.0625) { 3 };
\node[font=\tiny] at (0.125, 0.3125) { 3 };
\node[font=\tiny] at (0.125, 0.4375) { 9 };
\node[font=\tiny] at (0.125, 0.5625) { 3 };
\node[font=\tiny] at (0.125, 0.8125) { 3 };
\node[font=\tiny] at (0.1875, 0.125) { 5 };
\node[font=\tiny] at (0.1875, 0.25) { 5 };
\node[font=\tiny] at (0.25, 0.1875) { 11 };
\node[font=\tiny] at (0.25, 0.4375) { 7 };
\node[font=\tiny] at (0.1875, 0.625) { 5 };
\node[font=\tiny] at (0.1875, 0.75) { 5 };
\node[font=\tiny] at (0.25, 0.6875) { 11 };
\node[font=\tiny] at (0.3125, 0.0) { 3 };
\node[font=\tiny] at (0.375, 0.1875) { 13 };
\node[font=\tiny] at (0.3125, 0.375) { 3 };
\node[font=\tiny] at (0.3125, 0.5) { 3 };
\node[font=\tiny] at (0.375, 0.6875) { 13 };
\node[font=\tiny] at (0.3125, 0.875) { 3 };
\node[font=\tiny] at (0.4375, 0.0) { 63 };
\node[font=\tiny] at (0.4375, 0.125) { 61 };
\node[font=\tiny] at (0.4375, 0.25) { 35 };
\node[font=\tiny] at (0.5, 0.1875) { 13 };
\node[font=\tiny] at (0.4375, 0.375) { 33 };
\node[font=\tiny] at (0.4375, 0.5) { 31 };
\node[font=\tiny] at (0.4375, 0.625) { 29 };
\node[font=\tiny] at (0.4375, 0.75) { 3 };
\node[font=\tiny] at (0.5, 0.6875) { 13 };
\node[font=\tiny] at (0.5625, 0.0) { 3 };
\node[font=\tiny] at (0.625, 0.1875) { 11 };
\node[font=\tiny] at (0.5625, 0.375) { 3 };
\node[font=\tiny] at (0.5625, 0.5) { 3 };
\node[font=\tiny] at (0.625, 0.4375) { 7 };
\node[font=\tiny] at (0.625, 0.6875) { 11 };
\node[font=\tiny] at (0.5625, 0.875) { 3 };
\node[font=\tiny] at (0.6875, 0.125) { 5 };
\node[font=\tiny] at (0.75, 0.0625) { 3 };
\node[font=\tiny] at (0.6875, 0.25) { 5 };
\node[font=\tiny] at (0.75, 0.3125) { 3 };
\node[font=\tiny] at (0.75, 0.4375) { 9 };
\node[font=\tiny] at (0.6875, 0.625) { 5 };
\node[font=\tiny] at (0.75, 0.5625) { 3 };
\node[font=\tiny] at (0.6875, 0.75) { 5 };
\node[font=\tiny] at (0.75, 0.8125) { 3 };
\node[font=\tiny] at (0.875, 0.1875) { 3 };
\node[font=\tiny] at (0.875, 0.4375) { 11 };
\node[font=\tiny] at (0.875, 0.6875) { 3 };
		\end{tikzpicture}
		& \begin{tikzpicture}[scale = 2.8]
			\input{Figures/Curves/sfc_minicg_79.tex}
		\end{tikzpicture} \\
		\begin{tikzpicture}[scale = 2.8]
			\input{Figures/Curves/sfc_minicg_80.tex}
		\end{tikzpicture}
		& \begin{tikzpicture}[scale = 2.8]
			\draw[opacity = 0.2, scale=1/8, shift={(-1/2, -1/2)}] (0, 0) grid (8, 8);
\draw[opacity = 0.2, scale=1/8, shift={(-1/2, -1/2)}] (0, 0) grid (8, 8);
\fill[LimeGreen, opacity=0] (-0.0625, -0.0625) rectangle ++(0.125, 0.125);
\fill[LimeGreen, opacity=0.3333333333333333] (-0.0625, 0.0625) rectangle ++(0.125, 0.125);
\fill[LimeGreen, opacity=0.6666666666666666] (0.0625, 0.0625) rectangle ++(0.125, 0.125);
\fill[LimeGreen, opacity=0.1111111111111111] (0.0625, -0.0625) rectangle ++(0.125, 0.125);
\fill[LimeGreen, opacity=0.4444444444444444] (0.1875, -0.0625) rectangle ++(0.125, 0.125);
\fill[LimeGreen, opacity=0.7777777777777777] (0.3125, -0.0625) rectangle ++(0.125, 0.125);
\fill[LimeGreen, opacity=0.2222222222222222] (0.3125, 0.0625) rectangle ++(0.125, 0.125);
\fill[LimeGreen, opacity=0.5555555555555556] (0.1875, 0.0625) rectangle ++(0.125, 0.125);
\fill[LimeGreen, opacity=0.8888888888888888] (0.1875, 0.1875) rectangle ++(0.125, 0.125);
\fill[LimeGreen, opacity=0.037037037037037035] (0.3125, 0.1875) rectangle ++(0.125, 0.125);
\fill[LimeGreen, opacity=0.37037037037037035] (0.3125, 0.3125) rectangle ++(0.125, 0.125);
\fill[LimeGreen, opacity=0.7037037037037037] (0.1875, 0.3125) rectangle ++(0.125, 0.125);
\fill[LimeGreen, opacity=0.14814814814814814] (0.0625, 0.3125) rectangle ++(0.125, 0.125);
\fill[LimeGreen, opacity=0.48148148148148145] (0.0625, 0.1875) rectangle ++(0.125, 0.125);
\fill[LimeGreen, opacity=0.8148148148148147] (-0.0625, 0.1875) rectangle ++(0.125, 0.125);
\fill[LimeGreen, opacity=0.25925925925925924] (-0.0625, 0.3125) rectangle ++(0.125, 0.125);
\fill[LimeGreen, opacity=0.5925925925925926] (-0.0625, 0.4375) rectangle ++(0.125, 0.125);
\fill[LimeGreen, opacity=0.9259259259259258] (0.0625, 0.4375) rectangle ++(0.125, 0.125);
\fill[LimeGreen, opacity=0.07407407407407407] (0.0625, 0.5625) rectangle ++(0.125, 0.125);
\fill[LimeGreen, opacity=0.4074074074074074] (-0.0625, 0.5625) rectangle ++(0.125, 0.125);
\fill[LimeGreen, opacity=0.7407407407407407] (-0.0625, 0.6875) rectangle ++(0.125, 0.125);
\fill[LimeGreen, opacity=0.18518518518518517] (-0.0625, 0.8125) rectangle ++(0.125, 0.125);
\fill[LimeGreen, opacity=0.5185185185185185] (0.0625, 0.8125) rectangle ++(0.125, 0.125);
\fill[LimeGreen, opacity=0.8518518518518517] (0.0625, 0.6875) rectangle ++(0.125, 0.125);
\fill[LimeGreen, opacity=0.2962962962962963] (0.1875, 0.6875) rectangle ++(0.125, 0.125);
\fill[LimeGreen, opacity=0.6296296296296297] (0.1875, 0.8125) rectangle ++(0.125, 0.125);
\fill[LimeGreen, opacity=0.9629629629629629] (0.3125, 0.8125) rectangle ++(0.125, 0.125);
\fill[LimeGreen, opacity=0.012345679012345678] (0.3125, 0.6875) rectangle ++(0.125, 0.125);
\fill[LimeGreen, opacity=0.345679012345679] (0.3125, 0.5625) rectangle ++(0.125, 0.125);
\fill[LimeGreen, opacity=0.6790123456790123] (0.1875, 0.5625) rectangle ++(0.125, 0.125);
\fill[LimeGreen, opacity=0.12345679012345678] (0.1875, 0.4375) rectangle ++(0.125, 0.125);
\fill[LimeGreen, opacity=0.4567901234567901] (0.3125, 0.4375) rectangle ++(0.125, 0.125);
\fill[LimeGreen, opacity=0.7901234567901234] (0.4375, 0.4375) rectangle ++(0.125, 0.125);
\fill[LimeGreen, opacity=0.2345679012345679] (0.5625, 0.4375) rectangle ++(0.125, 0.125);
\fill[LimeGreen, opacity=0.5679012345679013] (0.5625, 0.5625) rectangle ++(0.125, 0.125);
\fill[LimeGreen, opacity=0.9012345679012346] (0.4375, 0.5625) rectangle ++(0.125, 0.125);
\fill[LimeGreen, opacity=0.04938271604938271] (0.4375, 0.6875) rectangle ++(0.125, 0.125);
\fill[LimeGreen, opacity=0.38271604938271603] (0.4375, 0.8125) rectangle ++(0.125, 0.125);
\fill[LimeGreen, opacity=0.7160493827160495] (0.5625, 0.8125) rectangle ++(0.125, 0.125);
\fill[LimeGreen, opacity=0.16049382716049382] (0.5625, 0.6875) rectangle ++(0.125, 0.125);
\fill[LimeGreen, opacity=0.49382716049382713] (0.6875, 0.6875) rectangle ++(0.125, 0.125);
\fill[LimeGreen, opacity=0.8271604938271604] (0.6875, 0.8125) rectangle ++(0.125, 0.125);
\fill[LimeGreen, opacity=0.2716049382716049] (0.8125, 0.8125) rectangle ++(0.125, 0.125);
\fill[LimeGreen, opacity=0.6049382716049383] (0.8125, 0.6875) rectangle ++(0.125, 0.125);
\fill[LimeGreen, opacity=0.9382716049382716] (0.8125, 0.5625) rectangle ++(0.125, 0.125);
\fill[LimeGreen, opacity=0.08641975308641975] (0.6875, 0.5625) rectangle ++(0.125, 0.125);
\fill[LimeGreen, opacity=0.41975308641975306] (0.6875, 0.4375) rectangle ++(0.125, 0.125);
\fill[LimeGreen, opacity=0.7530864197530864] (0.8125, 0.4375) rectangle ++(0.125, 0.125);
\fill[LimeGreen, opacity=0.19753086419753085] (0.8125, 0.3125) rectangle ++(0.125, 0.125);
\fill[LimeGreen, opacity=0.5308641975308641] (0.8125, 0.1875) rectangle ++(0.125, 0.125);
\fill[LimeGreen, opacity=0.8641975308641974] (0.6875, 0.1875) rectangle ++(0.125, 0.125);
\fill[LimeGreen, opacity=0.30864197530864196] (0.6875, 0.3125) rectangle ++(0.125, 0.125);
\fill[LimeGreen, opacity=0.6419753086419753] (0.5625, 0.3125) rectangle ++(0.125, 0.125);
\fill[LimeGreen, opacity=0.9753086419753085] (0.4375, 0.3125) rectangle ++(0.125, 0.125);
\fill[LimeGreen, opacity=0.024691358024691357] (0.4375, 0.1875) rectangle ++(0.125, 0.125);
\fill[LimeGreen, opacity=0.35802469135802467] (0.5625, 0.1875) rectangle ++(0.125, 0.125);
\fill[LimeGreen, opacity=0.691358024691358] (0.5625, 0.0625) rectangle ++(0.125, 0.125);
\fill[LimeGreen, opacity=0.13580246913580246] (0.4375, 0.0625) rectangle ++(0.125, 0.125);
\fill[LimeGreen, opacity=0.4691358024691358] (0.4375, -0.0625) rectangle ++(0.125, 0.125);
\fill[LimeGreen, opacity=0.802469135802469] (0.5625, -0.0625) rectangle ++(0.125, 0.125);
\fill[LimeGreen, opacity=0.24691358024691357] (0.6875, -0.0625) rectangle ++(0.125, 0.125);
\fill[LimeGreen, opacity=0.5802469135802469] (0.6875, 0.0625) rectangle ++(0.125, 0.125);
\fill[LimeGreen, opacity=0.9135802469135802] (0.8125, 0.0625) rectangle ++(0.125, 0.125);
\fill[LimeGreen, opacity=0.06172839506172839] (0.8125, -0.0625) rectangle ++(0.125, 0.125);
\draw[opacity=0.3] (0.0, 0.0)--(0.0, 0.125)--(0.125, 0.125)--(0.125, 0.0)--(0.25, 0.0)--(0.375, 0.0)--(0.375, 0.125)--(0.25, 0.125)--(0.25, 0.25)--(0.375, 0.25)--(0.375, 0.375)--(0.25, 0.375)--(0.125, 0.375)--(0.125, 0.25)--(0.0, 0.25)--(0.0, 0.375)--(0.0, 0.5)--(0.125, 0.5)--(0.125, 0.625)--(0.0, 0.625)--(0.0, 0.75)--(0.0, 0.875)--(0.125, 0.875)--(0.125, 0.75)--(0.25, 0.75)--(0.25, 0.875)--(0.375, 0.875)--(0.375, 0.75)--(0.375, 0.625)--(0.25, 0.625)--(0.25, 0.5)--(0.375, 0.5)--(0.5, 0.5)--(0.625, 0.5)--(0.625, 0.625)--(0.5, 0.625)--(0.5, 0.75)--(0.5, 0.875)--(0.625, 0.875)--(0.625, 0.75)--(0.75, 0.75)--(0.75, 0.875)--(0.875, 0.875)--(0.875, 0.75)--(0.875, 0.625)--(0.75, 0.625)--(0.75, 0.5)--(0.875, 0.5)--(0.875, 0.375)--(0.875, 0.25)--(0.75, 0.25)--(0.75, 0.375)--(0.625, 0.375)--(0.5, 0.375)--(0.5, 0.25)--(0.625, 0.25)--(0.625, 0.125)--(0.5, 0.125)--(0.5, 0.0)--(0.625, 0.0)--(0.75, 0.0)--(0.75, 0.125)--(0.875, 0.125)--(0.875, 0.0);
\node[font=\tiny] at (0.0, 0.1875) { 13 };
\node[font=\tiny] at (0.0, 0.5625) { 3 };
\node[font=\tiny] at (0.0625, 0.0) { 3 };
\node[font=\tiny] at (0.125, 0.1875) { 11 };
\node[font=\tiny] at (0.0625, 0.375) { 3 };
\node[font=\tiny] at (0.125, 0.4375) { 5 };
\node[font=\tiny] at (0.0625, 0.75) { 3 };
\node[font=\tiny] at (0.125, 0.6875) { 5 };
\node[font=\tiny] at (0.1875, 0.125) { 5 };
\node[font=\tiny] at (0.25, 0.0625) { 3 };
\node[font=\tiny] at (0.1875, 0.25) { 5 };
\node[font=\tiny] at (0.25, 0.3125) { 3 };
\node[font=\tiny] at (0.1875, 0.5) { 13 };
\node[font=\tiny] at (0.25, 0.4375) { 19 };
\node[font=\tiny] at (0.1875, 0.625) { 11 };
\node[font=\tiny] at (0.25, 0.6875) { 5 };
\node[font=\tiny] at (0.1875, 0.875) { 3 };
\node[font=\tiny] at (0.375, 0.1875) { 3 };
\node[font=\tiny] at (0.375, 0.4375) { 21 };
\node[font=\tiny] at (0.375, 0.5625) { 3 };
\node[font=\tiny] at (0.3125, 0.75) { 3 };
\node[font=\tiny] at (0.4375, 0.0) { 53 };
\node[font=\tiny] at (0.4375, 0.125) { 51 };
\node[font=\tiny] at (0.4375, 0.25) { 45 };
\node[font=\tiny] at (0.5, 0.1875) { 3 };
\node[font=\tiny] at (0.4375, 0.375) { 43 };
\node[font=\tiny] at (0.5, 0.4375) { 21 };
\node[font=\tiny] at (0.4375, 0.625) { 7 };
\node[font=\tiny] at (0.5, 0.5625) { 3 };
\node[font=\tiny] at (0.4375, 0.75) { 9 };
\node[font=\tiny] at (0.4375, 0.875) { 11 };
\node[font=\tiny] at (0.625, 0.0625) { 3 };
\node[font=\tiny] at (0.625, 0.3125) { 3 };
\node[font=\tiny] at (0.625, 0.4375) { 19 };
\node[font=\tiny] at (0.5625, 0.75) { 3 };
\node[font=\tiny] at (0.625, 0.6875) { 5 };
\node[font=\tiny] at (0.6875, 0.125) { 5 };
\node[font=\tiny] at (0.6875, 0.25) { 5 };
\node[font=\tiny] at (0.75, 0.1875) { 11 };
\node[font=\tiny] at (0.6875, 0.5) { 13 };
\node[font=\tiny] at (0.75, 0.4375) { 5 };
\node[font=\tiny] at (0.6875, 0.625) { 11 };
\node[font=\tiny] at (0.75, 0.6875) { 5 };
\node[font=\tiny] at (0.6875, 0.875) { 3 };
\node[font=\tiny] at (0.8125, 0.0) { 3 };
\node[font=\tiny] at (0.875, 0.1875) { 13 };
\node[font=\tiny] at (0.8125, 0.375) { 3 };
\node[font=\tiny] at (0.875, 0.5625) { 3 };
\node[font=\tiny] at (0.8125, 0.75) { 3 };
		\end{tikzpicture}
		& \begin{tikzpicture}[scale = 2.8]
			\draw[opacity = 0.2, scale=1/8, shift={(-1/2, -1/2)}] (0, 0) grid (8, 8);
\draw[opacity = 0.2, scale=1/8, shift={(-1/2, -1/2)}] (0, 0) grid (8, 8);
\fill[LimeGreen, opacity=0] (0.3125, -0.0625) rectangle ++(0.125, 0.125);
\fill[LimeGreen, opacity=0.3333333333333333] (0.3125, 0.0625) rectangle ++(0.125, 0.125);
\fill[LimeGreen, opacity=0.6666666666666666] (0.1875, 0.0625) rectangle ++(0.125, 0.125);
\fill[LimeGreen, opacity=0.1111111111111111] (0.1875, -0.0625) rectangle ++(0.125, 0.125);
\fill[LimeGreen, opacity=0.4444444444444444] (0.0625, -0.0625) rectangle ++(0.125, 0.125);
\fill[LimeGreen, opacity=0.7777777777777777] (-0.0625, -0.0625) rectangle ++(0.125, 0.125);
\fill[LimeGreen, opacity=0.2222222222222222] (-0.0625, 0.0625) rectangle ++(0.125, 0.125);
\fill[LimeGreen, opacity=0.5555555555555556] (0.0625, 0.0625) rectangle ++(0.125, 0.125);
\fill[LimeGreen, opacity=0.8888888888888888] (0.0625, 0.1875) rectangle ++(0.125, 0.125);
\fill[LimeGreen, opacity=0.037037037037037035] (-0.0625, 0.1875) rectangle ++(0.125, 0.125);
\fill[LimeGreen, opacity=0.37037037037037035] (-0.0625, 0.3125) rectangle ++(0.125, 0.125);
\fill[LimeGreen, opacity=0.7037037037037037] (0.0625, 0.3125) rectangle ++(0.125, 0.125);
\fill[LimeGreen, opacity=0.14814814814814814] (0.1875, 0.3125) rectangle ++(0.125, 0.125);
\fill[LimeGreen, opacity=0.48148148148148145] (0.1875, 0.1875) rectangle ++(0.125, 0.125);
\fill[LimeGreen, opacity=0.8148148148148147] (0.3125, 0.1875) rectangle ++(0.125, 0.125);
\fill[LimeGreen, opacity=0.25925925925925924] (0.3125, 0.3125) rectangle ++(0.125, 0.125);
\fill[LimeGreen, opacity=0.5925925925925926] (0.3125, 0.4375) rectangle ++(0.125, 0.125);
\fill[LimeGreen, opacity=0.9259259259259258] (0.3125, 0.5625) rectangle ++(0.125, 0.125);
\fill[LimeGreen, opacity=0.07407407407407407] (0.1875, 0.5625) rectangle ++(0.125, 0.125);
\fill[LimeGreen, opacity=0.4074074074074074] (0.1875, 0.4375) rectangle ++(0.125, 0.125);
\fill[LimeGreen, opacity=0.7407407407407407] (0.0625, 0.4375) rectangle ++(0.125, 0.125);
\fill[LimeGreen, opacity=0.18518518518518517] (-0.0625, 0.4375) rectangle ++(0.125, 0.125);
\fill[LimeGreen, opacity=0.5185185185185185] (-0.0625, 0.5625) rectangle ++(0.125, 0.125);
\fill[LimeGreen, opacity=0.8518518518518517] (0.0625, 0.5625) rectangle ++(0.125, 0.125);
\fill[LimeGreen, opacity=0.2962962962962963] (0.0625, 0.6875) rectangle ++(0.125, 0.125);
\fill[LimeGreen, opacity=0.6296296296296297] (-0.0625, 0.6875) rectangle ++(0.125, 0.125);
\fill[LimeGreen, opacity=0.9629629629629629] (-0.0625, 0.8125) rectangle ++(0.125, 0.125);
\fill[LimeGreen, opacity=0.012345679012345678] (0.0625, 0.8125) rectangle ++(0.125, 0.125);
\fill[LimeGreen, opacity=0.345679012345679] (0.1875, 0.8125) rectangle ++(0.125, 0.125);
\fill[LimeGreen, opacity=0.6790123456790123] (0.1875, 0.6875) rectangle ++(0.125, 0.125);
\fill[LimeGreen, opacity=0.12345679012345678] (0.3125, 0.6875) rectangle ++(0.125, 0.125);
\fill[LimeGreen, opacity=0.4567901234567901] (0.3125, 0.8125) rectangle ++(0.125, 0.125);
\fill[LimeGreen, opacity=0.7901234567901234] (0.4375, 0.8125) rectangle ++(0.125, 0.125);
\fill[LimeGreen, opacity=0.2345679012345679] (0.4375, 0.6875) rectangle ++(0.125, 0.125);
\fill[LimeGreen, opacity=0.5679012345679013] (0.5625, 0.6875) rectangle ++(0.125, 0.125);
\fill[LimeGreen, opacity=0.9012345679012346] (0.5625, 0.8125) rectangle ++(0.125, 0.125);
\fill[LimeGreen, opacity=0.04938271604938271] (0.6875, 0.8125) rectangle ++(0.125, 0.125);
\fill[LimeGreen, opacity=0.38271604938271603] (0.8125, 0.8125) rectangle ++(0.125, 0.125);
\fill[LimeGreen, opacity=0.7160493827160495] (0.8125, 0.6875) rectangle ++(0.125, 0.125);
\fill[LimeGreen, opacity=0.16049382716049382] (0.6875, 0.6875) rectangle ++(0.125, 0.125);
\fill[LimeGreen, opacity=0.49382716049382713] (0.6875, 0.5625) rectangle ++(0.125, 0.125);
\fill[LimeGreen, opacity=0.8271604938271604] (0.8125, 0.5625) rectangle ++(0.125, 0.125);
\fill[LimeGreen, opacity=0.2716049382716049] (0.8125, 0.4375) rectangle ++(0.125, 0.125);
\fill[LimeGreen, opacity=0.6049382716049383] (0.6875, 0.4375) rectangle ++(0.125, 0.125);
\fill[LimeGreen, opacity=0.9382716049382716] (0.5625, 0.4375) rectangle ++(0.125, 0.125);
\fill[LimeGreen, opacity=0.08641975308641975] (0.5625, 0.5625) rectangle ++(0.125, 0.125);
\fill[LimeGreen, opacity=0.41975308641975306] (0.4375, 0.5625) rectangle ++(0.125, 0.125);
\fill[LimeGreen, opacity=0.7530864197530864] (0.4375, 0.4375) rectangle ++(0.125, 0.125);
\fill[LimeGreen, opacity=0.19753086419753085] (0.4375, 0.3125) rectangle ++(0.125, 0.125);
\fill[LimeGreen, opacity=0.5308641975308641] (0.4375, 0.1875) rectangle ++(0.125, 0.125);
\fill[LimeGreen, opacity=0.8641975308641974] (0.5625, 0.1875) rectangle ++(0.125, 0.125);
\fill[LimeGreen, opacity=0.30864197530864196] (0.5625, 0.3125) rectangle ++(0.125, 0.125);
\fill[LimeGreen, opacity=0.6419753086419753] (0.6875, 0.3125) rectangle ++(0.125, 0.125);
\fill[LimeGreen, opacity=0.9753086419753085] (0.8125, 0.3125) rectangle ++(0.125, 0.125);
\fill[LimeGreen, opacity=0.024691358024691357] (0.8125, 0.1875) rectangle ++(0.125, 0.125);
\fill[LimeGreen, opacity=0.35802469135802467] (0.6875, 0.1875) rectangle ++(0.125, 0.125);
\fill[LimeGreen, opacity=0.691358024691358] (0.6875, 0.0625) rectangle ++(0.125, 0.125);
\fill[LimeGreen, opacity=0.13580246913580246] (0.8125, 0.0625) rectangle ++(0.125, 0.125);
\fill[LimeGreen, opacity=0.4691358024691358] (0.8125, -0.0625) rectangle ++(0.125, 0.125);
\fill[LimeGreen, opacity=0.802469135802469] (0.6875, -0.0625) rectangle ++(0.125, 0.125);
\fill[LimeGreen, opacity=0.24691358024691357] (0.5625, -0.0625) rectangle ++(0.125, 0.125);
\fill[LimeGreen, opacity=0.5802469135802469] (0.5625, 0.0625) rectangle ++(0.125, 0.125);
\fill[LimeGreen, opacity=0.9135802469135802] (0.4375, 0.0625) rectangle ++(0.125, 0.125);
\fill[LimeGreen, opacity=0.06172839506172839] (0.4375, -0.0625) rectangle ++(0.125, 0.125);
\draw[opacity=0.3] (0.375, 0.0)--(0.375, 0.125)--(0.25, 0.125)--(0.25, 0.0)--(0.125, 0.0)--(0.0, 0.0)--(0.0, 0.125)--(0.125, 0.125)--(0.125, 0.25)--(0.0, 0.25)--(0.0, 0.375)--(0.125, 0.375)--(0.25, 0.375)--(0.25, 0.25)--(0.375, 0.25)--(0.375, 0.375)--(0.375, 0.5)--(0.375, 0.625)--(0.25, 0.625)--(0.25, 0.5)--(0.125, 0.5)--(0.0, 0.5)--(0.0, 0.625)--(0.125, 0.625)--(0.125, 0.75)--(0.0, 0.75)--(0.0, 0.875)--(0.125, 0.875)--(0.25, 0.875)--(0.25, 0.75)--(0.375, 0.75)--(0.375, 0.875)--(0.5, 0.875)--(0.5, 0.75)--(0.625, 0.75)--(0.625, 0.875)--(0.75, 0.875)--(0.875, 0.875)--(0.875, 0.75)--(0.75, 0.75)--(0.75, 0.625)--(0.875, 0.625)--(0.875, 0.5)--(0.75, 0.5)--(0.625, 0.5)--(0.625, 0.625)--(0.5, 0.625)--(0.5, 0.5)--(0.5, 0.375)--(0.5, 0.25)--(0.625, 0.25)--(0.625, 0.375)--(0.75, 0.375)--(0.875, 0.375)--(0.875, 0.25)--(0.75, 0.25)--(0.75, 0.125)--(0.875, 0.125)--(0.875, 0.0)--(0.75, 0.0)--(0.625, 0.0)--(0.625, 0.125)--(0.5, 0.125)--(0.5, 0.0);
\node[font=\tiny] at (0.0, 0.1875) { 3 };
\node[font=\tiny] at (0.0, 0.4375) { 11 };
\node[font=\tiny] at (0.0, 0.6875) { 3 };
\node[font=\tiny] at (0.125, 0.0625) { 3 };
\node[font=\tiny] at (0.125, 0.3125) { 3 };
\node[font=\tiny] at (0.125, 0.4375) { 9 };
\node[font=\tiny] at (0.125, 0.5625) { 3 };
\node[font=\tiny] at (0.125, 0.8125) { 3 };
\node[font=\tiny] at (0.1875, 0.125) { 5 };
\node[font=\tiny] at (0.1875, 0.25) { 5 };
\node[font=\tiny] at (0.25, 0.1875) { 11 };
\node[font=\tiny] at (0.25, 0.4375) { 7 };
\node[font=\tiny] at (0.1875, 0.625) { 5 };
\node[font=\tiny] at (0.1875, 0.75) { 5 };
\node[font=\tiny] at (0.25, 0.6875) { 11 };
\node[font=\tiny] at (0.3125, 0.0) { 3 };
\node[font=\tiny] at (0.375, 0.1875) { 13 };
\node[font=\tiny] at (0.3125, 0.375) { 3 };
\node[font=\tiny] at (0.3125, 0.5) { 3 };
\node[font=\tiny] at (0.375, 0.6875) { 13 };
\node[font=\tiny] at (0.3125, 0.875) { 3 };
\node[font=\tiny] at (0.4375, 0.0) { 63 };
\node[font=\tiny] at (0.4375, 0.125) { 61 };
\node[font=\tiny] at (0.4375, 0.25) { 35 };
\node[font=\tiny] at (0.5, 0.1875) { 13 };
\node[font=\tiny] at (0.4375, 0.375) { 33 };
\node[font=\tiny] at (0.4375, 0.5) { 31 };
\node[font=\tiny] at (0.4375, 0.625) { 29 };
\node[font=\tiny] at (0.4375, 0.75) { 3 };
\node[font=\tiny] at (0.5, 0.6875) { 13 };
\node[font=\tiny] at (0.5625, 0.0) { 3 };
\node[font=\tiny] at (0.625, 0.1875) { 11 };
\node[font=\tiny] at (0.5625, 0.375) { 3 };
\node[font=\tiny] at (0.5625, 0.5) { 3 };
\node[font=\tiny] at (0.625, 0.4375) { 7 };
\node[font=\tiny] at (0.625, 0.6875) { 11 };
\node[font=\tiny] at (0.5625, 0.875) { 3 };
\node[font=\tiny] at (0.6875, 0.125) { 5 };
\node[font=\tiny] at (0.75, 0.0625) { 3 };
\node[font=\tiny] at (0.6875, 0.25) { 5 };
\node[font=\tiny] at (0.75, 0.3125) { 3 };
\node[font=\tiny] at (0.75, 0.4375) { 9 };
\node[font=\tiny] at (0.6875, 0.625) { 5 };
\node[font=\tiny] at (0.75, 0.5625) { 3 };
\node[font=\tiny] at (0.6875, 0.75) { 5 };
\node[font=\tiny] at (0.75, 0.8125) { 3 };
\node[font=\tiny] at (0.875, 0.1875) { 3 };
\node[font=\tiny] at (0.875, 0.4375) { 11 };
\node[font=\tiny] at (0.875, 0.6875) { 3 };
		\end{tikzpicture}
		& \begin{tikzpicture}[scale = 2.8]
			\input{Figures/Curves/sfc_minicg_83.tex}
		\end{tikzpicture} \\
		\begin{tikzpicture}[scale = 2.8]
			\draw[opacity = 0.2, scale=1/8, shift={(-1/2, -1/2)}] (0, 0) grid (8, 8);
\draw[opacity = 0.2, scale=1/8, shift={(-1/2, -1/2)}] (0, 0) grid (8, 8);
\fill[LimeGreen, opacity=0] (-0.0625, -0.0625) rectangle ++(0.125, 0.125);
\fill[LimeGreen, opacity=0.2] (0.0625, -0.0625) rectangle ++(0.125, 0.125);
\fill[LimeGreen, opacity=0.4] (-0.0625, 0.0625) rectangle ++(0.125, 0.125);
\fill[LimeGreen, opacity=0.6000000000000001] (0.0625, 0.0625) rectangle ++(0.125, 0.125);
\fill[LimeGreen, opacity=0.8] (0.1875, -0.0625) rectangle ++(0.125, 0.125);
\fill[LimeGreen, opacity=0.04] (0.3125, -0.0625) rectangle ++(0.125, 0.125);
\fill[LimeGreen, opacity=0.24000000000000002] (0.1875, 0.0625) rectangle ++(0.125, 0.125);
\fill[LimeGreen, opacity=0.44] (0.3125, 0.0625) rectangle ++(0.125, 0.125);
\fill[LimeGreen, opacity=0.6400000000000001] (-0.0625, 0.1875) rectangle ++(0.125, 0.125);
\fill[LimeGreen, opacity=0.8400000000000001] (0.0625, 0.1875) rectangle ++(0.125, 0.125);
\fill[LimeGreen, opacity=0.08] (-0.0625, 0.3125) rectangle ++(0.125, 0.125);
\fill[LimeGreen, opacity=0.28] (0.0625, 0.3125) rectangle ++(0.125, 0.125);
\fill[LimeGreen, opacity=0.48000000000000004] (0.1875, 0.1875) rectangle ++(0.125, 0.125);
\fill[LimeGreen, opacity=0.68] (0.3125, 0.1875) rectangle ++(0.125, 0.125);
\fill[LimeGreen, opacity=0.88] (0.1875, 0.3125) rectangle ++(0.125, 0.125);
\fill[LimeGreen, opacity=0.12] (0.3125, 0.3125) rectangle ++(0.125, 0.125);
\fill[LimeGreen, opacity=0.32] (0.4375, -0.0625) rectangle ++(0.125, 0.125);
\fill[LimeGreen, opacity=0.52] (0.5625, -0.0625) rectangle ++(0.125, 0.125);
\fill[LimeGreen, opacity=0.7200000000000001] (0.4375, 0.0625) rectangle ++(0.125, 0.125);
\fill[LimeGreen, opacity=0.92] (0.5625, 0.0625) rectangle ++(0.125, 0.125);
\fill[LimeGreen, opacity=0.16] (0.6875, -0.0625) rectangle ++(0.125, 0.125);
\fill[LimeGreen, opacity=0.36] (0.8125, -0.0625) rectangle ++(0.125, 0.125);
\fill[LimeGreen, opacity=0.56] (0.6875, 0.0625) rectangle ++(0.125, 0.125);
\fill[LimeGreen, opacity=0.7600000000000001] (0.8125, 0.0625) rectangle ++(0.125, 0.125);
\fill[LimeGreen, opacity=0.9600000000000001] (0.4375, 0.1875) rectangle ++(0.125, 0.125);
\fill[LimeGreen, opacity=0.008] (0.5625, 0.1875) rectangle ++(0.125, 0.125);
\fill[LimeGreen, opacity=0.20800000000000002] (0.4375, 0.3125) rectangle ++(0.125, 0.125);
\fill[LimeGreen, opacity=0.40800000000000003] (0.5625, 0.3125) rectangle ++(0.125, 0.125);
\fill[LimeGreen, opacity=0.6080000000000001] (0.6875, 0.1875) rectangle ++(0.125, 0.125);
\fill[LimeGreen, opacity=0.808] (0.8125, 0.1875) rectangle ++(0.125, 0.125);
\fill[LimeGreen, opacity=0.048] (0.6875, 0.3125) rectangle ++(0.125, 0.125);
\fill[LimeGreen, opacity=0.24800000000000003] (0.8125, 0.3125) rectangle ++(0.125, 0.125);
\fill[LimeGreen, opacity=0.448] (-0.0625, 0.4375) rectangle ++(0.125, 0.125);
\fill[LimeGreen, opacity=0.6480000000000001] (0.0625, 0.4375) rectangle ++(0.125, 0.125);
\fill[LimeGreen, opacity=0.8480000000000001] (-0.0625, 0.5625) rectangle ++(0.125, 0.125);
\fill[LimeGreen, opacity=0.088] (0.0625, 0.5625) rectangle ++(0.125, 0.125);
\fill[LimeGreen, opacity=0.28800000000000003] (0.1875, 0.4375) rectangle ++(0.125, 0.125);
\fill[LimeGreen, opacity=0.48800000000000004] (0.3125, 0.4375) rectangle ++(0.125, 0.125);
\fill[LimeGreen, opacity=0.6880000000000001] (0.1875, 0.5625) rectangle ++(0.125, 0.125);
\fill[LimeGreen, opacity=0.888] (0.3125, 0.5625) rectangle ++(0.125, 0.125);
\fill[LimeGreen, opacity=0.128] (-0.0625, 0.6875) rectangle ++(0.125, 0.125);
\fill[LimeGreen, opacity=0.328] (0.0625, 0.6875) rectangle ++(0.125, 0.125);
\fill[LimeGreen, opacity=0.528] (-0.0625, 0.8125) rectangle ++(0.125, 0.125);
\fill[LimeGreen, opacity=0.7280000000000001] (0.0625, 0.8125) rectangle ++(0.125, 0.125);
\fill[LimeGreen, opacity=0.928] (0.1875, 0.6875) rectangle ++(0.125, 0.125);
\fill[LimeGreen, opacity=0.168] (0.3125, 0.6875) rectangle ++(0.125, 0.125);
\fill[LimeGreen, opacity=0.368] (0.1875, 0.8125) rectangle ++(0.125, 0.125);
\fill[LimeGreen, opacity=0.5680000000000001] (0.3125, 0.8125) rectangle ++(0.125, 0.125);
\fill[LimeGreen, opacity=0.7680000000000001] (0.4375, 0.4375) rectangle ++(0.125, 0.125);
\fill[LimeGreen, opacity=0.9680000000000001] (0.5625, 0.4375) rectangle ++(0.125, 0.125);
\fill[LimeGreen, opacity=0.016] (0.4375, 0.5625) rectangle ++(0.125, 0.125);
\fill[LimeGreen, opacity=0.21600000000000003] (0.5625, 0.5625) rectangle ++(0.125, 0.125);
\fill[LimeGreen, opacity=0.41600000000000004] (0.6875, 0.4375) rectangle ++(0.125, 0.125);
\fill[LimeGreen, opacity=0.6160000000000001] (0.8125, 0.4375) rectangle ++(0.125, 0.125);
\fill[LimeGreen, opacity=0.8160000000000001] (0.6875, 0.5625) rectangle ++(0.125, 0.125);
\fill[LimeGreen, opacity=0.056] (0.8125, 0.5625) rectangle ++(0.125, 0.125);
\fill[LimeGreen, opacity=0.256] (0.4375, 0.6875) rectangle ++(0.125, 0.125);
\fill[LimeGreen, opacity=0.456] (0.5625, 0.6875) rectangle ++(0.125, 0.125);
\fill[LimeGreen, opacity=0.6560000000000001] (0.4375, 0.8125) rectangle ++(0.125, 0.125);
\fill[LimeGreen, opacity=0.8560000000000001] (0.5625, 0.8125) rectangle ++(0.125, 0.125);
\fill[LimeGreen, opacity=0.096] (0.6875, 0.6875) rectangle ++(0.125, 0.125);
\fill[LimeGreen, opacity=0.29600000000000004] (0.8125, 0.6875) rectangle ++(0.125, 0.125);
\fill[LimeGreen, opacity=0.49600000000000005] (0.6875, 0.8125) rectangle ++(0.125, 0.125);
\fill[LimeGreen, opacity=0.6960000000000001] (0.8125, 0.8125) rectangle ++(0.125, 0.125);
\draw[opacity=0.3] (0.0, 0.0)--(0.125, 0.0)--(0.0, 0.125)--(0.125, 0.125)--(0.25, 0.0)--(0.375, 0.0)--(0.25, 0.125)--(0.375, 0.125)--(0.0, 0.25)--(0.125, 0.25)--(0.0, 0.375)--(0.125, 0.375)--(0.25, 0.25)--(0.375, 0.25)--(0.25, 0.375)--(0.375, 0.375)--(0.5, 0.0)--(0.625, 0.0)--(0.5, 0.125)--(0.625, 0.125)--(0.75, 0.0)--(0.875, 0.0)--(0.75, 0.125)--(0.875, 0.125)--(0.5, 0.25)--(0.625, 0.25)--(0.5, 0.375)--(0.625, 0.375)--(0.75, 0.25)--(0.875, 0.25)--(0.75, 0.375)--(0.875, 0.375)--(0.0, 0.5)--(0.125, 0.5)--(0.0, 0.625)--(0.125, 0.625)--(0.25, 0.5)--(0.375, 0.5)--(0.25, 0.625)--(0.375, 0.625)--(0.0, 0.75)--(0.125, 0.75)--(0.0, 0.875)--(0.125, 0.875)--(0.25, 0.75)--(0.375, 0.75)--(0.25, 0.875)--(0.375, 0.875)--(0.5, 0.5)--(0.625, 0.5)--(0.5, 0.625)--(0.625, 0.625)--(0.75, 0.5)--(0.875, 0.5)--(0.75, 0.625)--(0.875, 0.625)--(0.5, 0.75)--(0.625, 0.75)--(0.5, 0.875)--(0.625, 0.875)--(0.75, 0.75)--(0.875, 0.75)--(0.75, 0.875)--(0.875, 0.875);
\node[font=\tiny] at (0.0, 0.0625) { 2 };
\node[font=\tiny] at (0.0, 0.1875) { 6 };
\node[font=\tiny] at (0.0, 0.3125) { 2 };
\node[font=\tiny] at (0.0, 0.4375) { 22 };
\node[font=\tiny] at (0.0, 0.5625) { 2 };
\node[font=\tiny] at (0.0, 0.6875) { 6 };
\node[font=\tiny] at (0.0, 0.8125) { 2 };
\node[font=\tiny] at (0.125, 0.0625) { 2 };
\node[font=\tiny] at (0.125, 0.1875) { 6 };
\node[font=\tiny] at (0.125, 0.3125) { 2 };
\node[font=\tiny] at (0.125, 0.4375) { 22 };
\node[font=\tiny] at (0.125, 0.5625) { 2 };
\node[font=\tiny] at (0.125, 0.6875) { 6 };
\node[font=\tiny] at (0.125, 0.8125) { 2 };
\node[font=\tiny] at (0.1875, 0.0) { 3 };
\node[font=\tiny] at (0.1875, 0.125) { 3 };
\node[font=\tiny] at (0.25, 0.0625) { 2 };
\node[font=\tiny] at (0.1875, 0.25) { 3 };
\node[font=\tiny] at (0.25, 0.1875) { 6 };
\node[font=\tiny] at (0.1875, 0.375) { 3 };
\node[font=\tiny] at (0.25, 0.3125) { 2 };
\node[font=\tiny] at (0.1875, 0.5) { 3 };
\node[font=\tiny] at (0.25, 0.4375) { 22 };
\node[font=\tiny] at (0.1875, 0.625) { 3 };
\node[font=\tiny] at (0.25, 0.5625) { 2 };
\node[font=\tiny] at (0.1875, 0.75) { 3 };
\node[font=\tiny] at (0.25, 0.6875) { 6 };
\node[font=\tiny] at (0.1875, 0.875) { 3 };
\node[font=\tiny] at (0.25, 0.8125) { 2 };
\node[font=\tiny] at (0.375, 0.0625) { 2 };
\node[font=\tiny] at (0.375, 0.1875) { 6 };
\node[font=\tiny] at (0.375, 0.3125) { 2 };
\node[font=\tiny] at (0.375, 0.4375) { 22 };
\node[font=\tiny] at (0.375, 0.5625) { 2 };
\node[font=\tiny] at (0.375, 0.6875) { 6 };
\node[font=\tiny] at (0.375, 0.8125) { 2 };
\node[font=\tiny] at (0.4375, 0.0) { 11 };
\node[font=\tiny] at (0.4375, 0.125) { 11 };
\node[font=\tiny] at (0.5, 0.0625) { 2 };
\node[font=\tiny] at (0.4375, 0.25) { 11 };
\node[font=\tiny] at (0.5, 0.1875) { 6 };
\node[font=\tiny] at (0.4375, 0.375) { 11 };
\node[font=\tiny] at (0.5, 0.3125) { 2 };
\node[font=\tiny] at (0.4375, 0.5) { 11 };
\node[font=\tiny] at (0.5, 0.4375) { 22 };
\node[font=\tiny] at (0.4375, 0.625) { 11 };
\node[font=\tiny] at (0.5, 0.5625) { 2 };
\node[font=\tiny] at (0.4375, 0.75) { 11 };
\node[font=\tiny] at (0.5, 0.6875) { 6 };
\node[font=\tiny] at (0.4375, 0.875) { 11 };
\node[font=\tiny] at (0.5, 0.8125) { 2 };
\node[font=\tiny] at (0.625, 0.0625) { 2 };
\node[font=\tiny] at (0.625, 0.1875) { 6 };
\node[font=\tiny] at (0.625, 0.3125) { 2 };
\node[font=\tiny] at (0.625, 0.4375) { 22 };
\node[font=\tiny] at (0.625, 0.5625) { 2 };
\node[font=\tiny] at (0.625, 0.6875) { 6 };
\node[font=\tiny] at (0.625, 0.8125) { 2 };
\node[font=\tiny] at (0.6875, 0.0) { 3 };
\node[font=\tiny] at (0.6875, 0.125) { 3 };
\node[font=\tiny] at (0.75, 0.0625) { 2 };
\node[font=\tiny] at (0.6875, 0.25) { 3 };
\node[font=\tiny] at (0.75, 0.1875) { 6 };
\node[font=\tiny] at (0.6875, 0.375) { 3 };
\node[font=\tiny] at (0.75, 0.3125) { 2 };
\node[font=\tiny] at (0.6875, 0.5) { 3 };
\node[font=\tiny] at (0.75, 0.4375) { 22 };
\node[font=\tiny] at (0.6875, 0.625) { 3 };
\node[font=\tiny] at (0.75, 0.5625) { 2 };
\node[font=\tiny] at (0.6875, 0.75) { 3 };
\node[font=\tiny] at (0.75, 0.6875) { 6 };
\node[font=\tiny] at (0.6875, 0.875) { 3 };
\node[font=\tiny] at (0.75, 0.8125) { 2 };
\node[font=\tiny] at (0.875, 0.0625) { 2 };
\node[font=\tiny] at (0.875, 0.1875) { 6 };
\node[font=\tiny] at (0.875, 0.3125) { 2 };
\node[font=\tiny] at (0.875, 0.4375) { 22 };
\node[font=\tiny] at (0.875, 0.5625) { 2 };
\node[font=\tiny] at (0.875, 0.6875) { 6 };
\node[font=\tiny] at (0.875, 0.8125) { 2 };
		\end{tikzpicture}
		& \begin{tikzpicture}[scale = 2.8]
			\draw[opacity = 0.2, scale=1/8, shift={(-1/2, -1/2)}] (0, 0) grid (8, 8);
\draw[opacity = 0.2, scale=1/8, shift={(-1/2, -1/2)}] (0, 0) grid (8, 8);
\fill[LimeGreen, opacity=0] (-0.0625, -0.0625) rectangle ++(0.125, 0.125);
\fill[LimeGreen, opacity=0.2] (-0.0625, 0.0625) rectangle ++(0.125, 0.125);
\fill[LimeGreen, opacity=0.4] (0.0625, 0.0625) rectangle ++(0.125, 0.125);
\fill[LimeGreen, opacity=0.6000000000000001] (0.0625, -0.0625) rectangle ++(0.125, 0.125);
\fill[LimeGreen, opacity=0.8] (0.1875, -0.0625) rectangle ++(0.125, 0.125);
\fill[LimeGreen, opacity=0.04] (0.3125, -0.0625) rectangle ++(0.125, 0.125);
\fill[LimeGreen, opacity=0.24000000000000002] (0.3125, 0.0625) rectangle ++(0.125, 0.125);
\fill[LimeGreen, opacity=0.44] (0.1875, 0.0625) rectangle ++(0.125, 0.125);
\fill[LimeGreen, opacity=0.6400000000000001] (0.1875, 0.1875) rectangle ++(0.125, 0.125);
\fill[LimeGreen, opacity=0.8400000000000001] (0.3125, 0.1875) rectangle ++(0.125, 0.125);
\fill[LimeGreen, opacity=0.08] (0.3125, 0.3125) rectangle ++(0.125, 0.125);
\fill[LimeGreen, opacity=0.28] (0.1875, 0.3125) rectangle ++(0.125, 0.125);
\fill[LimeGreen, opacity=0.48000000000000004] (0.0625, 0.3125) rectangle ++(0.125, 0.125);
\fill[LimeGreen, opacity=0.68] (0.0625, 0.1875) rectangle ++(0.125, 0.125);
\fill[LimeGreen, opacity=0.88] (-0.0625, 0.1875) rectangle ++(0.125, 0.125);
\fill[LimeGreen, opacity=0.12] (-0.0625, 0.3125) rectangle ++(0.125, 0.125);
\fill[LimeGreen, opacity=0.32] (-0.0625, 0.4375) rectangle ++(0.125, 0.125);
\fill[LimeGreen, opacity=0.52] (0.0625, 0.4375) rectangle ++(0.125, 0.125);
\fill[LimeGreen, opacity=0.7200000000000001] (0.0625, 0.5625) rectangle ++(0.125, 0.125);
\fill[LimeGreen, opacity=0.92] (-0.0625, 0.5625) rectangle ++(0.125, 0.125);
\fill[LimeGreen, opacity=0.16] (-0.0625, 0.6875) rectangle ++(0.125, 0.125);
\fill[LimeGreen, opacity=0.36] (-0.0625, 0.8125) rectangle ++(0.125, 0.125);
\fill[LimeGreen, opacity=0.56] (0.0625, 0.8125) rectangle ++(0.125, 0.125);
\fill[LimeGreen, opacity=0.7600000000000001] (0.0625, 0.6875) rectangle ++(0.125, 0.125);
\fill[LimeGreen, opacity=0.9600000000000001] (0.1875, 0.6875) rectangle ++(0.125, 0.125);
\fill[LimeGreen, opacity=0.008] (0.1875, 0.8125) rectangle ++(0.125, 0.125);
\fill[LimeGreen, opacity=0.20800000000000002] (0.3125, 0.8125) rectangle ++(0.125, 0.125);
\fill[LimeGreen, opacity=0.40800000000000003] (0.3125, 0.6875) rectangle ++(0.125, 0.125);
\fill[LimeGreen, opacity=0.6080000000000001] (0.3125, 0.5625) rectangle ++(0.125, 0.125);
\fill[LimeGreen, opacity=0.808] (0.1875, 0.5625) rectangle ++(0.125, 0.125);
\fill[LimeGreen, opacity=0.048] (0.1875, 0.4375) rectangle ++(0.125, 0.125);
\fill[LimeGreen, opacity=0.24800000000000003] (0.3125, 0.4375) rectangle ++(0.125, 0.125);
\fill[LimeGreen, opacity=0.448] (0.4375, 0.4375) rectangle ++(0.125, 0.125);
\fill[LimeGreen, opacity=0.6480000000000001] (0.5625, 0.4375) rectangle ++(0.125, 0.125);
\fill[LimeGreen, opacity=0.8480000000000001] (0.5625, 0.5625) rectangle ++(0.125, 0.125);
\fill[LimeGreen, opacity=0.088] (0.4375, 0.5625) rectangle ++(0.125, 0.125);
\fill[LimeGreen, opacity=0.28800000000000003] (0.4375, 0.6875) rectangle ++(0.125, 0.125);
\fill[LimeGreen, opacity=0.48800000000000004] (0.4375, 0.8125) rectangle ++(0.125, 0.125);
\fill[LimeGreen, opacity=0.6880000000000001] (0.5625, 0.8125) rectangle ++(0.125, 0.125);
\fill[LimeGreen, opacity=0.888] (0.5625, 0.6875) rectangle ++(0.125, 0.125);
\fill[LimeGreen, opacity=0.128] (0.6875, 0.6875) rectangle ++(0.125, 0.125);
\fill[LimeGreen, opacity=0.328] (0.6875, 0.8125) rectangle ++(0.125, 0.125);
\fill[LimeGreen, opacity=0.528] (0.8125, 0.8125) rectangle ++(0.125, 0.125);
\fill[LimeGreen, opacity=0.7280000000000001] (0.8125, 0.6875) rectangle ++(0.125, 0.125);
\fill[LimeGreen, opacity=0.928] (0.8125, 0.5625) rectangle ++(0.125, 0.125);
\fill[LimeGreen, opacity=0.168] (0.6875, 0.5625) rectangle ++(0.125, 0.125);
\fill[LimeGreen, opacity=0.368] (0.6875, 0.4375) rectangle ++(0.125, 0.125);
\fill[LimeGreen, opacity=0.5680000000000001] (0.8125, 0.4375) rectangle ++(0.125, 0.125);
\fill[LimeGreen, opacity=0.7680000000000001] (0.8125, 0.3125) rectangle ++(0.125, 0.125);
\fill[LimeGreen, opacity=0.9680000000000001] (0.8125, 0.1875) rectangle ++(0.125, 0.125);
\fill[LimeGreen, opacity=0.016] (0.6875, 0.1875) rectangle ++(0.125, 0.125);
\fill[LimeGreen, opacity=0.21600000000000003] (0.6875, 0.3125) rectangle ++(0.125, 0.125);
\fill[LimeGreen, opacity=0.41600000000000004] (0.5625, 0.3125) rectangle ++(0.125, 0.125);
\fill[LimeGreen, opacity=0.6160000000000001] (0.4375, 0.3125) rectangle ++(0.125, 0.125);
\fill[LimeGreen, opacity=0.8160000000000001] (0.4375, 0.1875) rectangle ++(0.125, 0.125);
\fill[LimeGreen, opacity=0.056] (0.5625, 0.1875) rectangle ++(0.125, 0.125);
\fill[LimeGreen, opacity=0.256] (0.5625, 0.0625) rectangle ++(0.125, 0.125);
\fill[LimeGreen, opacity=0.456] (0.4375, 0.0625) rectangle ++(0.125, 0.125);
\fill[LimeGreen, opacity=0.6560000000000001] (0.4375, -0.0625) rectangle ++(0.125, 0.125);
\fill[LimeGreen, opacity=0.8560000000000001] (0.5625, -0.0625) rectangle ++(0.125, 0.125);
\fill[LimeGreen, opacity=0.096] (0.6875, -0.0625) rectangle ++(0.125, 0.125);
\fill[LimeGreen, opacity=0.29600000000000004] (0.6875, 0.0625) rectangle ++(0.125, 0.125);
\fill[LimeGreen, opacity=0.49600000000000005] (0.8125, 0.0625) rectangle ++(0.125, 0.125);
\fill[LimeGreen, opacity=0.6960000000000001] (0.8125, -0.0625) rectangle ++(0.125, 0.125);
\draw[opacity=0.3] (0.0, 0.0)--(0.0, 0.125)--(0.125, 0.125)--(0.125, 0.0)--(0.25, 0.0)--(0.375, 0.0)--(0.375, 0.125)--(0.25, 0.125)--(0.25, 0.25)--(0.375, 0.25)--(0.375, 0.375)--(0.25, 0.375)--(0.125, 0.375)--(0.125, 0.25)--(0.0, 0.25)--(0.0, 0.375)--(0.0, 0.5)--(0.125, 0.5)--(0.125, 0.625)--(0.0, 0.625)--(0.0, 0.75)--(0.0, 0.875)--(0.125, 0.875)--(0.125, 0.75)--(0.25, 0.75)--(0.25, 0.875)--(0.375, 0.875)--(0.375, 0.75)--(0.375, 0.625)--(0.25, 0.625)--(0.25, 0.5)--(0.375, 0.5)--(0.5, 0.5)--(0.625, 0.5)--(0.625, 0.625)--(0.5, 0.625)--(0.5, 0.75)--(0.5, 0.875)--(0.625, 0.875)--(0.625, 0.75)--(0.75, 0.75)--(0.75, 0.875)--(0.875, 0.875)--(0.875, 0.75)--(0.875, 0.625)--(0.75, 0.625)--(0.75, 0.5)--(0.875, 0.5)--(0.875, 0.375)--(0.875, 0.25)--(0.75, 0.25)--(0.75, 0.375)--(0.625, 0.375)--(0.5, 0.375)--(0.5, 0.25)--(0.625, 0.25)--(0.625, 0.125)--(0.5, 0.125)--(0.5, 0.0)--(0.625, 0.0)--(0.75, 0.0)--(0.75, 0.125)--(0.875, 0.125)--(0.875, 0.0);
\node[font=\tiny] at (0.0, 0.1875) { 13 };
\node[font=\tiny] at (0.0, 0.5625) { 3 };
\node[font=\tiny] at (0.0625, 0.0) { 3 };
\node[font=\tiny] at (0.125, 0.1875) { 11 };
\node[font=\tiny] at (0.0625, 0.375) { 3 };
\node[font=\tiny] at (0.125, 0.4375) { 5 };
\node[font=\tiny] at (0.0625, 0.75) { 3 };
\node[font=\tiny] at (0.125, 0.6875) { 5 };
\node[font=\tiny] at (0.1875, 0.125) { 5 };
\node[font=\tiny] at (0.25, 0.0625) { 3 };
\node[font=\tiny] at (0.1875, 0.25) { 5 };
\node[font=\tiny] at (0.25, 0.3125) { 3 };
\node[font=\tiny] at (0.1875, 0.5) { 13 };
\node[font=\tiny] at (0.25, 0.4375) { 19 };
\node[font=\tiny] at (0.1875, 0.625) { 11 };
\node[font=\tiny] at (0.25, 0.6875) { 5 };
\node[font=\tiny] at (0.1875, 0.875) { 3 };
\node[font=\tiny] at (0.375, 0.1875) { 3 };
\node[font=\tiny] at (0.375, 0.4375) { 21 };
\node[font=\tiny] at (0.375, 0.5625) { 3 };
\node[font=\tiny] at (0.3125, 0.75) { 3 };
\node[font=\tiny] at (0.4375, 0.0) { 53 };
\node[font=\tiny] at (0.4375, 0.125) { 51 };
\node[font=\tiny] at (0.4375, 0.25) { 45 };
\node[font=\tiny] at (0.5, 0.1875) { 3 };
\node[font=\tiny] at (0.4375, 0.375) { 43 };
\node[font=\tiny] at (0.5, 0.4375) { 21 };
\node[font=\tiny] at (0.4375, 0.625) { 7 };
\node[font=\tiny] at (0.5, 0.5625) { 3 };
\node[font=\tiny] at (0.4375, 0.75) { 9 };
\node[font=\tiny] at (0.4375, 0.875) { 11 };
\node[font=\tiny] at (0.625, 0.0625) { 3 };
\node[font=\tiny] at (0.625, 0.3125) { 3 };
\node[font=\tiny] at (0.625, 0.4375) { 19 };
\node[font=\tiny] at (0.5625, 0.75) { 3 };
\node[font=\tiny] at (0.625, 0.6875) { 5 };
\node[font=\tiny] at (0.6875, 0.125) { 5 };
\node[font=\tiny] at (0.6875, 0.25) { 5 };
\node[font=\tiny] at (0.75, 0.1875) { 11 };
\node[font=\tiny] at (0.6875, 0.5) { 13 };
\node[font=\tiny] at (0.75, 0.4375) { 5 };
\node[font=\tiny] at (0.6875, 0.625) { 11 };
\node[font=\tiny] at (0.75, 0.6875) { 5 };
\node[font=\tiny] at (0.6875, 0.875) { 3 };
\node[font=\tiny] at (0.8125, 0.0) { 3 };
\node[font=\tiny] at (0.875, 0.1875) { 13 };
\node[font=\tiny] at (0.8125, 0.375) { 3 };
\node[font=\tiny] at (0.875, 0.5625) { 3 };
\node[font=\tiny] at (0.8125, 0.75) { 3 };
		\end{tikzpicture}
		& \begin{tikzpicture}[scale = 2.8]
			\draw[opacity = 0.2, scale=1/8, shift={(-1/2, -1/2)}] (0, 0) grid (8, 8);
\draw[opacity = 0.2, scale=1/8, shift={(-1/2, -1/2)}] (0, 0) grid (8, 8);
\fill[LimeGreen, opacity=0] (0.3125, -0.0625) rectangle ++(0.125, 0.125);
\fill[LimeGreen, opacity=0.2] (0.3125, 0.0625) rectangle ++(0.125, 0.125);
\fill[LimeGreen, opacity=0.4] (0.1875, 0.0625) rectangle ++(0.125, 0.125);
\fill[LimeGreen, opacity=0.6000000000000001] (0.1875, -0.0625) rectangle ++(0.125, 0.125);
\fill[LimeGreen, opacity=0.8] (0.0625, -0.0625) rectangle ++(0.125, 0.125);
\fill[LimeGreen, opacity=0.04] (-0.0625, -0.0625) rectangle ++(0.125, 0.125);
\fill[LimeGreen, opacity=0.24000000000000002] (-0.0625, 0.0625) rectangle ++(0.125, 0.125);
\fill[LimeGreen, opacity=0.44] (0.0625, 0.0625) rectangle ++(0.125, 0.125);
\fill[LimeGreen, opacity=0.6400000000000001] (0.0625, 0.1875) rectangle ++(0.125, 0.125);
\fill[LimeGreen, opacity=0.8400000000000001] (-0.0625, 0.1875) rectangle ++(0.125, 0.125);
\fill[LimeGreen, opacity=0.08] (-0.0625, 0.3125) rectangle ++(0.125, 0.125);
\fill[LimeGreen, opacity=0.28] (0.0625, 0.3125) rectangle ++(0.125, 0.125);
\fill[LimeGreen, opacity=0.48000000000000004] (0.1875, 0.3125) rectangle ++(0.125, 0.125);
\fill[LimeGreen, opacity=0.68] (0.1875, 0.1875) rectangle ++(0.125, 0.125);
\fill[LimeGreen, opacity=0.88] (0.3125, 0.1875) rectangle ++(0.125, 0.125);
\fill[LimeGreen, opacity=0.12] (0.3125, 0.3125) rectangle ++(0.125, 0.125);
\fill[LimeGreen, opacity=0.32] (0.3125, 0.4375) rectangle ++(0.125, 0.125);
\fill[LimeGreen, opacity=0.52] (0.3125, 0.5625) rectangle ++(0.125, 0.125);
\fill[LimeGreen, opacity=0.7200000000000001] (0.1875, 0.5625) rectangle ++(0.125, 0.125);
\fill[LimeGreen, opacity=0.92] (0.1875, 0.4375) rectangle ++(0.125, 0.125);
\fill[LimeGreen, opacity=0.16] (0.0625, 0.4375) rectangle ++(0.125, 0.125);
\fill[LimeGreen, opacity=0.36] (-0.0625, 0.4375) rectangle ++(0.125, 0.125);
\fill[LimeGreen, opacity=0.56] (-0.0625, 0.5625) rectangle ++(0.125, 0.125);
\fill[LimeGreen, opacity=0.7600000000000001] (0.0625, 0.5625) rectangle ++(0.125, 0.125);
\fill[LimeGreen, opacity=0.9600000000000001] (0.0625, 0.6875) rectangle ++(0.125, 0.125);
\fill[LimeGreen, opacity=0.008] (-0.0625, 0.6875) rectangle ++(0.125, 0.125);
\fill[LimeGreen, opacity=0.20800000000000002] (-0.0625, 0.8125) rectangle ++(0.125, 0.125);
\fill[LimeGreen, opacity=0.40800000000000003] (0.0625, 0.8125) rectangle ++(0.125, 0.125);
\fill[LimeGreen, opacity=0.6080000000000001] (0.1875, 0.8125) rectangle ++(0.125, 0.125);
\fill[LimeGreen, opacity=0.808] (0.1875, 0.6875) rectangle ++(0.125, 0.125);
\fill[LimeGreen, opacity=0.048] (0.3125, 0.6875) rectangle ++(0.125, 0.125);
\fill[LimeGreen, opacity=0.24800000000000003] (0.3125, 0.8125) rectangle ++(0.125, 0.125);
\fill[LimeGreen, opacity=0.448] (0.4375, 0.8125) rectangle ++(0.125, 0.125);
\fill[LimeGreen, opacity=0.6480000000000001] (0.4375, 0.6875) rectangle ++(0.125, 0.125);
\fill[LimeGreen, opacity=0.8480000000000001] (0.5625, 0.6875) rectangle ++(0.125, 0.125);
\fill[LimeGreen, opacity=0.088] (0.5625, 0.8125) rectangle ++(0.125, 0.125);
\fill[LimeGreen, opacity=0.28800000000000003] (0.6875, 0.8125) rectangle ++(0.125, 0.125);
\fill[LimeGreen, opacity=0.48800000000000004] (0.8125, 0.8125) rectangle ++(0.125, 0.125);
\fill[LimeGreen, opacity=0.6880000000000001] (0.8125, 0.6875) rectangle ++(0.125, 0.125);
\fill[LimeGreen, opacity=0.888] (0.6875, 0.6875) rectangle ++(0.125, 0.125);
\fill[LimeGreen, opacity=0.128] (0.6875, 0.5625) rectangle ++(0.125, 0.125);
\fill[LimeGreen, opacity=0.328] (0.8125, 0.5625) rectangle ++(0.125, 0.125);
\fill[LimeGreen, opacity=0.528] (0.8125, 0.4375) rectangle ++(0.125, 0.125);
\fill[LimeGreen, opacity=0.7280000000000001] (0.6875, 0.4375) rectangle ++(0.125, 0.125);
\fill[LimeGreen, opacity=0.928] (0.5625, 0.4375) rectangle ++(0.125, 0.125);
\fill[LimeGreen, opacity=0.168] (0.5625, 0.5625) rectangle ++(0.125, 0.125);
\fill[LimeGreen, opacity=0.368] (0.4375, 0.5625) rectangle ++(0.125, 0.125);
\fill[LimeGreen, opacity=0.5680000000000001] (0.4375, 0.4375) rectangle ++(0.125, 0.125);
\fill[LimeGreen, opacity=0.7680000000000001] (0.4375, 0.3125) rectangle ++(0.125, 0.125);
\fill[LimeGreen, opacity=0.9680000000000001] (0.4375, 0.1875) rectangle ++(0.125, 0.125);
\fill[LimeGreen, opacity=0.016] (0.5625, 0.1875) rectangle ++(0.125, 0.125);
\fill[LimeGreen, opacity=0.21600000000000003] (0.5625, 0.3125) rectangle ++(0.125, 0.125);
\fill[LimeGreen, opacity=0.41600000000000004] (0.6875, 0.3125) rectangle ++(0.125, 0.125);
\fill[LimeGreen, opacity=0.6160000000000001] (0.8125, 0.3125) rectangle ++(0.125, 0.125);
\fill[LimeGreen, opacity=0.8160000000000001] (0.8125, 0.1875) rectangle ++(0.125, 0.125);
\fill[LimeGreen, opacity=0.056] (0.6875, 0.1875) rectangle ++(0.125, 0.125);
\fill[LimeGreen, opacity=0.256] (0.6875, 0.0625) rectangle ++(0.125, 0.125);
\fill[LimeGreen, opacity=0.456] (0.8125, 0.0625) rectangle ++(0.125, 0.125);
\fill[LimeGreen, opacity=0.6560000000000001] (0.8125, -0.0625) rectangle ++(0.125, 0.125);
\fill[LimeGreen, opacity=0.8560000000000001] (0.6875, -0.0625) rectangle ++(0.125, 0.125);
\fill[LimeGreen, opacity=0.096] (0.5625, -0.0625) rectangle ++(0.125, 0.125);
\fill[LimeGreen, opacity=0.29600000000000004] (0.5625, 0.0625) rectangle ++(0.125, 0.125);
\fill[LimeGreen, opacity=0.49600000000000005] (0.4375, 0.0625) rectangle ++(0.125, 0.125);
\fill[LimeGreen, opacity=0.6960000000000001] (0.4375, -0.0625) rectangle ++(0.125, 0.125);
\draw[opacity=0.3] (0.375, 0.0)--(0.375, 0.125)--(0.25, 0.125)--(0.25, 0.0)--(0.125, 0.0)--(0.0, 0.0)--(0.0, 0.125)--(0.125, 0.125)--(0.125, 0.25)--(0.0, 0.25)--(0.0, 0.375)--(0.125, 0.375)--(0.25, 0.375)--(0.25, 0.25)--(0.375, 0.25)--(0.375, 0.375)--(0.375, 0.5)--(0.375, 0.625)--(0.25, 0.625)--(0.25, 0.5)--(0.125, 0.5)--(0.0, 0.5)--(0.0, 0.625)--(0.125, 0.625)--(0.125, 0.75)--(0.0, 0.75)--(0.0, 0.875)--(0.125, 0.875)--(0.25, 0.875)--(0.25, 0.75)--(0.375, 0.75)--(0.375, 0.875)--(0.5, 0.875)--(0.5, 0.75)--(0.625, 0.75)--(0.625, 0.875)--(0.75, 0.875)--(0.875, 0.875)--(0.875, 0.75)--(0.75, 0.75)--(0.75, 0.625)--(0.875, 0.625)--(0.875, 0.5)--(0.75, 0.5)--(0.625, 0.5)--(0.625, 0.625)--(0.5, 0.625)--(0.5, 0.5)--(0.5, 0.375)--(0.5, 0.25)--(0.625, 0.25)--(0.625, 0.375)--(0.75, 0.375)--(0.875, 0.375)--(0.875, 0.25)--(0.75, 0.25)--(0.75, 0.125)--(0.875, 0.125)--(0.875, 0.0)--(0.75, 0.0)--(0.625, 0.0)--(0.625, 0.125)--(0.5, 0.125)--(0.5, 0.0);
\node[font=\tiny] at (0.0, 0.1875) { 3 };
\node[font=\tiny] at (0.0, 0.4375) { 11 };
\node[font=\tiny] at (0.0, 0.6875) { 3 };
\node[font=\tiny] at (0.125, 0.0625) { 3 };
\node[font=\tiny] at (0.125, 0.3125) { 3 };
\node[font=\tiny] at (0.125, 0.4375) { 9 };
\node[font=\tiny] at (0.125, 0.5625) { 3 };
\node[font=\tiny] at (0.125, 0.8125) { 3 };
\node[font=\tiny] at (0.1875, 0.125) { 5 };
\node[font=\tiny] at (0.1875, 0.25) { 5 };
\node[font=\tiny] at (0.25, 0.1875) { 11 };
\node[font=\tiny] at (0.25, 0.4375) { 7 };
\node[font=\tiny] at (0.1875, 0.625) { 5 };
\node[font=\tiny] at (0.1875, 0.75) { 5 };
\node[font=\tiny] at (0.25, 0.6875) { 11 };
\node[font=\tiny] at (0.3125, 0.0) { 3 };
\node[font=\tiny] at (0.375, 0.1875) { 13 };
\node[font=\tiny] at (0.3125, 0.375) { 3 };
\node[font=\tiny] at (0.3125, 0.5) { 3 };
\node[font=\tiny] at (0.375, 0.6875) { 13 };
\node[font=\tiny] at (0.3125, 0.875) { 3 };
\node[font=\tiny] at (0.4375, 0.0) { 63 };
\node[font=\tiny] at (0.4375, 0.125) { 61 };
\node[font=\tiny] at (0.4375, 0.25) { 35 };
\node[font=\tiny] at (0.5, 0.1875) { 13 };
\node[font=\tiny] at (0.4375, 0.375) { 33 };
\node[font=\tiny] at (0.4375, 0.5) { 31 };
\node[font=\tiny] at (0.4375, 0.625) { 29 };
\node[font=\tiny] at (0.4375, 0.75) { 3 };
\node[font=\tiny] at (0.5, 0.6875) { 13 };
\node[font=\tiny] at (0.5625, 0.0) { 3 };
\node[font=\tiny] at (0.625, 0.1875) { 11 };
\node[font=\tiny] at (0.5625, 0.375) { 3 };
\node[font=\tiny] at (0.5625, 0.5) { 3 };
\node[font=\tiny] at (0.625, 0.4375) { 7 };
\node[font=\tiny] at (0.625, 0.6875) { 11 };
\node[font=\tiny] at (0.5625, 0.875) { 3 };
\node[font=\tiny] at (0.6875, 0.125) { 5 };
\node[font=\tiny] at (0.75, 0.0625) { 3 };
\node[font=\tiny] at (0.6875, 0.25) { 5 };
\node[font=\tiny] at (0.75, 0.3125) { 3 };
\node[font=\tiny] at (0.75, 0.4375) { 9 };
\node[font=\tiny] at (0.6875, 0.625) { 5 };
\node[font=\tiny] at (0.75, 0.5625) { 3 };
\node[font=\tiny] at (0.6875, 0.75) { 5 };
\node[font=\tiny] at (0.75, 0.8125) { 3 };
\node[font=\tiny] at (0.875, 0.1875) { 3 };
\node[font=\tiny] at (0.875, 0.4375) { 11 };
\node[font=\tiny] at (0.875, 0.6875) { 3 };
		\end{tikzpicture}
		& \begin{tikzpicture}[scale = 2.8]
			\input{Figures/Curves/sfc_minicg_87.tex}
		\end{tikzpicture} \\
  Morton & Hilbert & Moore & Peano
  \end{tabular}
  \caption{The numbers on the edges shared by neighboring pixels are
  the difference of pixel indices enumerated along the depicted space-filling curve.
  Along the curve, this difference is one and implicitly represented by the imprinted curves.
  Diagrams in a column belong to the named space-filling curve. From top to bottom,
  the rows show the differences for the radical inverses $\phi_2$, $\phi_3$, and $\phi_5$,
  while the shade of each pixel represents the
  value of the radical inverse of the pixel index. As can be seen, the differences contain
  symmetries and repetitive patterns, which result in visible correlations in the
  pixels shades. For curves in base $b = 2$, it is easy to spot a
  checker-boarding effect. For the Peano curve, symmetries along the diagonal may be
  observed.
  Besides this illustration of the principle,
  structures may be more visible at higher resolutions, see Fig.~\ref{Fig:Dithermap}}.
  \label{Fig:Correlation}
\end{figure}

\begin{figure}
  \centering
  \begin{tabular}{cccc}
  \includegraphics[width=0.24\linewidth]{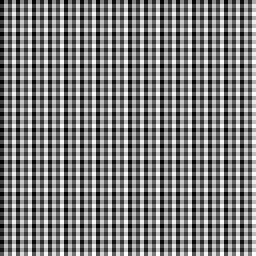}
  & \includegraphics[width=0.24\linewidth]{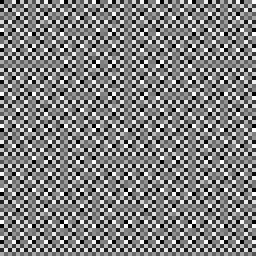}
  & \includegraphics[width=0.24\linewidth]{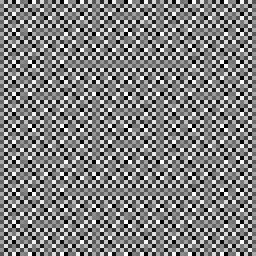}
  & \includegraphics[width=0.24\linewidth]{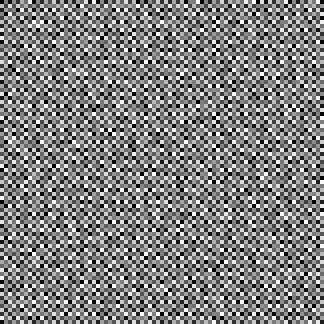} \\
  \includegraphics[width=0.24\linewidth]{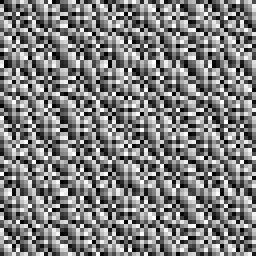}
  & \includegraphics[width=0.24\linewidth]{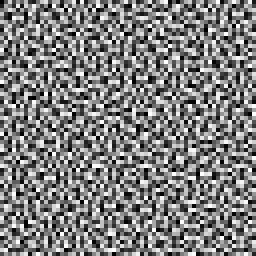}
  & \includegraphics[width=0.24\linewidth]{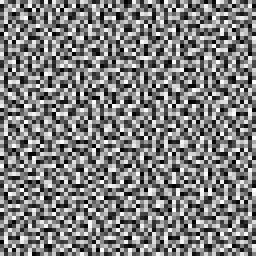}
  & \includegraphics[width=0.24\linewidth]{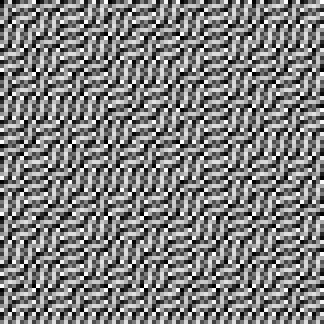} \\
  \includegraphics[width=0.24\linewidth]{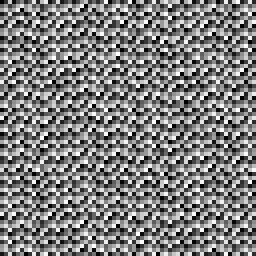}
  & \includegraphics[width=0.24\linewidth]{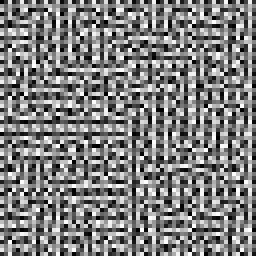}
  & \includegraphics[width=0.24\linewidth]{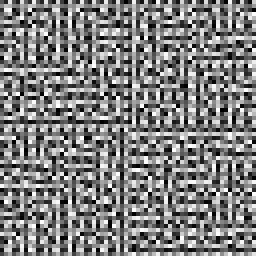}
  & \includegraphics[width=0.24\linewidth]{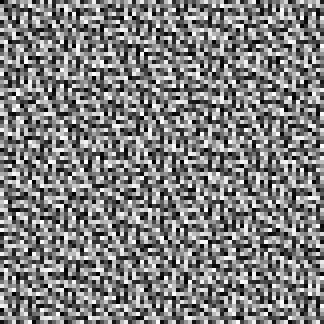} \\
  \includegraphics[width=0.24\linewidth]{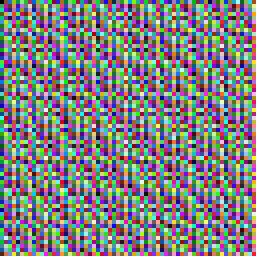}
  & \includegraphics[width=0.24\linewidth]{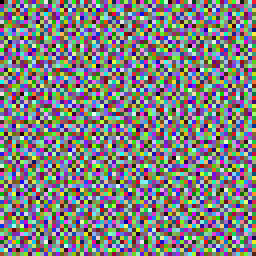}
  & \includegraphics[width=0.24\linewidth]{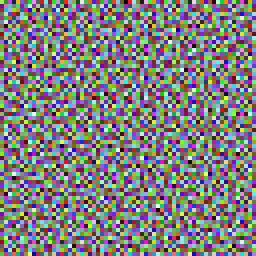}
  & \includegraphics[width=0.24\linewidth]{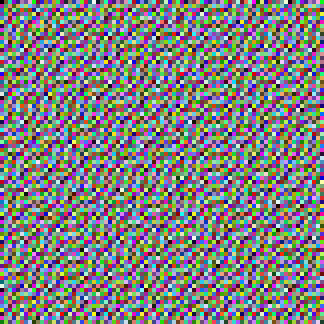} \\
  Morton & Hilbert & Moore & Peano
  \end{tabular}
  \caption{From top to bottom, the rows display the values of the
  radical inverses $\phi_2$, $\phi_3$, and $\phi_5$
  as gray values, and all previous three superimposed
  by assigning them to  the RGB channels of an image.
  The columns indicate the space filling curve used for enumeration.
  The resolution for the Morton, Hilbert, and Moore
  curves in base $b = 2$ is $64 \times 64$ pixels, while
  for the Peano curve in base $b = 3$ we display $81 \times 81$
  pixels. 
  For the eye it is easy to spot regular structures that are caused by correlations
  between the single low discrepancy sequences and the space filling curves.
  Superimposing them as in the bottom row,
  it becomes harder to identify the correlations with the Moore and Hilbert
  curves. 
  }
  \label{Fig:Dithermap}
\end{figure}

\begin{figure}
  \centering
  \begin{tabular}{cccc}
    \includegraphics[width=0.24\linewidth]{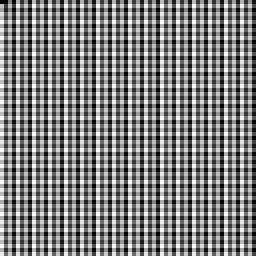}&
    \includegraphics[width=0.24\linewidth]{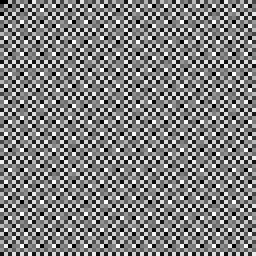}&
    \includegraphics[width=0.24\linewidth]{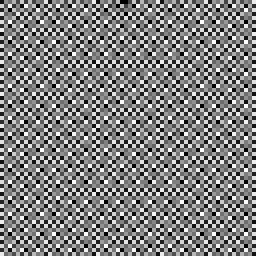}&
    \includegraphics[width=0.24\linewidth]{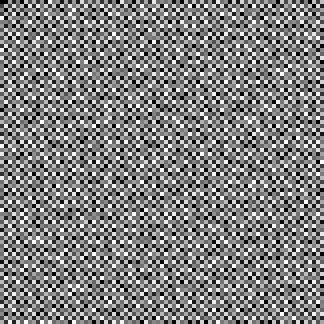}\\
    \includegraphics[width=0.24\linewidth]{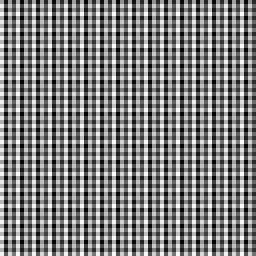}&
    \includegraphics[width=0.24\linewidth]{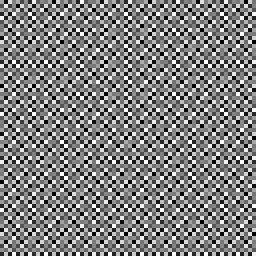}&
    \includegraphics[width=0.24\linewidth]{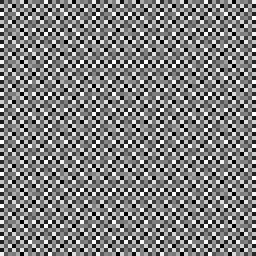}&
    \includegraphics[width=0.24\linewidth]{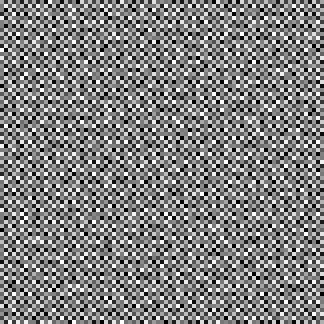}\\
  \includegraphics[width=0.24\linewidth]{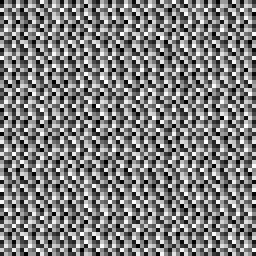}
  & \includegraphics[width=0.24\linewidth]{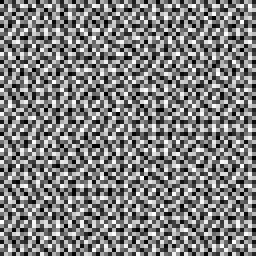}
  & \includegraphics[width=0.24\linewidth]{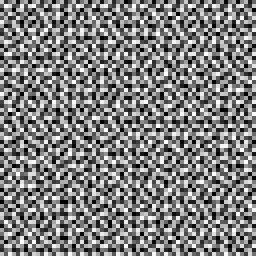}
  & \includegraphics[width=0.24\linewidth]{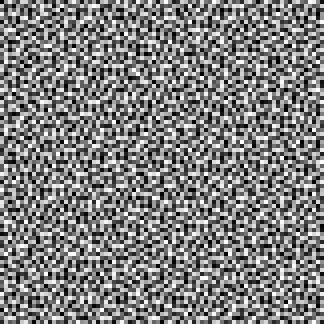} \\
    Morton & Hilbert & Moore & Peano
  \end{tabular}
  \caption{While scrambling may attenuate visible correlations between pixels,
  it cannot completely remove the correlations. 
  The first row shows $\phi_2$ with random digit scrambling and the second
  row shows $\phi_2$  with Owen scrambling. The bottom row shows
  $\phi_5$ improved according to Faure~\cite{ImprovedHalton}.
  As compared to Fig.~\ref{Fig:Dithermap}, scrambling $\phi_2$ does not at all
  help when used with the Morton curve and still leaves some visible lines
  indicative of the quadrant structure of the Hilbert and Moore curves. For
  the example of $\phi_5$, scrambling helps most for the correlations when using the
  Peano curve but does not dramatically attenuate the artifacts when using
  the other space filling curves. 
  }
  \label{Fig:Scrambling}
\end{figure}

\subsection{Correlation in Space-Filling Curves} \label{Sec:Correlation}

We show that the correlation patterns visualized in Fig.~\ref{Fig:Correlation} emanate
from enumerating radical inverses \cite{Nie:92} along a space-filling curve. A radical inverse
\begin{eqnarray}
  \phi_b: \mathbb{N}_0 & \rightarrow & \mathbb{Q} \cap [0,1) \nonumber\\
  i  =  \sum_{k = 0}^\infty a_k(i) b^k
   & \mapsto & \sum_{k = 0}^\infty a_k(i) b^{-k - 1}  \label{Eqn:RadInv}
\end{eqnarray}
maps a non-negative integer to the unit interval by reflecting
its digits $a_k(i)$ in base $b$ at the decimal point. The Halton
sequence is an example of an infinite-dimensional low discrepancy
sequence. Each dimension is a radical inverse, where all the
bases are relatively co-prime. The uniformity of the simple construction can be
improved by applying a permutation to the $k$-th digit $a_k(i)$ of the index $i$ represented in
base $b$ before radical inversion. Zaremba \cite{Zar:70} has been successful with the simple permutation
$\pi_b(a_k(i)) := (a_k(i) + k) \bmod b$, while later on Faure \cite{Faure:92}
developed a more general set of permutations improving upon
Zaremba's results.

Now taking a look at the Morton curve in Fig.~\ref{Fig:Correlation}, it becomes
obvious that the Morton index is either odd or even per column of pixels. As a
consequence, computing the radical inverse $\phi_2$ of this index, which amounts to
bits reversal of the index, results in $\phi_2 < \frac{1}{2}$
in even columns and $\phi_2 \geq \frac{1}{2}$ in odd columns \cite[Sect.3.1]{Z-Sampler}.
Similarly, the second least significant bit of the Morton index is on and off along the rows of pixels.
This correlation results in striping artifacts along rows and columns of pixels,
especially visible at one sample per pixel as shown in Fig.~\ref{Fig:Dithermap}.

Other space-filling curves expose correlations, too, the reason being simply that
the differences of indices of pixels are deterministic
and correlated. Fig.~\ref{Fig:Correlation} shows the index differences
larger than one for neighboring pixels. 
While the Hilbert and Moore curves are in base $b = 2$, they expose
correlations with $\phi_3$ and higher bases. For example, stripes are visible
along lines of differences that are a multiple of the radical inverses' base 3. Similarly,
the Hilbert curve has many adjacent pairs of pixels, whose
difference of indices is 3. Radical inverses in these pixels in base $b = 3$ hence
are correlated. The larger such clusters, the more prominent is the visible artifact.

While correlations are to be expected when the base of the low discrepancy
sequence and the space filling curve are not co-prime,
correlation structures may become visible 
whenever the differences of the pixel indices along
a space filling curve are correlated to the base of the low discrepancy
sequence in a regular way.

Owen scrambling \cite{Owen:1995} may resolve these correlations, because it recursively
partitions the unit interval and randomly swaps the partitions independently. For the case of the Morton curve, Ahmed \cite{Z-Sampler} developed a
scrambling scheme, where it is sufficient to apply the recursive random swapping
procedure to the index of the pixel along the Morton curve. The algorithm amounts to applying
random permutations to the contiguous block of indices belonging to each quadrant along
the hierarchy of the Morton curve.

As exemplified in Fig.~\ref{Fig:Scrambling}, even recursive scrambling cannot
remove all correlation artifacts. For example, in base $b = 2$, the recursive
structure of scrambling correlates with the block structure of the Morton, Moore,
and Hilbert curves. Yet, visible correlation artifacts are attenuated. Note that
for the most significant bit both digit scrambling and Owen scrambling are identical.
No matter how or whether this bit is scrambled, consecutive pixels along a space filling
curve are hence larger and less or equal to $\frac{1}{2}$. In contrast to the Morton
curve, this guarantees a good uniformity of values in the neighborhood of each pixel
for the Hilbert, Moore, and Peano curves.
For the example of base $b= 5$, scrambling may attenuate the visible structures
and yet cannot resolve the correlations of the base of the radical inverse
and the difference of pixel indices as shown in Fig.~\ref{Fig:Correlation}.

\subsection{Blue-Noise Dithered Sampling} \label{Sec:BlueNoiseMasks}

While some correlations are visible in low dimensions and at low sampling rates,
sampling light transport paths requires many more dimensions. Fig.~\ref{Fig:Dithermap}
illustrates that already overlaying the first three radical inverses as RGB values
hides most of the disturbing artifacts.

The maps in Fig.~\ref{Fig:Dithermap} resemble the maps
used for blue-noise dithered sampling \cite{DitheredSampling} and may be used for the same
purposes. As opposed to the optimization process required to create blue-noise dither maps,
enumerating low discrepancy sequences along a space-filling curve allows one to approximate
the desired spectral properties by just selecting components without the restriction to
low dimension and without the need to store lookup tables. This approach is partially explored
in Sect.~\ref{Sec:DeterministicShift}.

\section{Progressive Image Synthesis} \label{Sec:Progressive}

Progressive image synthesis
continues sampling within a given sample or time budget or terminates
sampling once a selected image quality has been reached \cite{Iray_report}.
In what follows, we discuss extensions of the consistent algorithm
of the previous section to enable adaptive sampling.

\subsection{Deterministic Cranley-Patterson Rotation} \label{Sec:DeterministicShift}

Similar to the algorithm detailed in Sect.~\ref{Sec:HilbertEnumeration}, a low
discrepancy sequence is enumerated along the Hilbert curve at one sample
per pixel. The vector of the low discrepancy sequence assigned to a pixel
then is used to perform a deterministic Cranley-Patterson rotation \cite{Cranley:76}.
This way, the same sequence of samples may be used across all pixels,
however, shifted individually.

The Cranley-Patterson rotation is implemented as component-wise
addition modulo one. Yet, using one and the same low discrepancy
sequence for both shifting and sampling may expose visible correlation
artifacts at low sampling rates.
This is the case when enumerating the improved Halton sequence
\cite{ImprovedHalton} along the Hilbert curve to shift the same
sequence per pixel. While Cranley-Patterson rotations work with
any point set, they work best with a point set designed for the unit torus
such as rank-1 lattices and rank-1 lattice
sequences \cite{eLattices:01}. For results, see Fig.~\ref{Fig:FiniteSamples}.

\subsection{Randomization} \label{Sec:ProgressiveRand}

Array-RQMC algorithms \cite{SortingArrayQMC} randomize the low discrepancy sequence
for each iteration. This approach is straightforward to apply to the
algorithm in Sect.~\ref{Sec:HilbertEnumeration}: for each pass, the low
discrepancy sequence is randomized and the results are accumulated until
the termination by an empirical error criterion is triggered or a given time budget expired.
The repeated randomization will eventually average out the correlation artifacts in the
rendered frame.

While randomizing the low discrepancy sequence and accumulating results is
unbiased and allows for unbiased variance estimation, some uniformity and hence
convergence speed is sacrificed. Therefore, we aim at a deterministic and
consistent algorithm, which in addition is simpler to execute and reproduce on massively
parallel computer systems \cite{NutshellQMC,Iray_report}.

\subsection{Contiguous Segments of one Low Discrepancy Sequence}

Progressive sampling may be implemented by iterating the algorithm
in Sect.~\ref{Sec:HilbertEnumeration}. To increase the uniformity
of the samples in a pixel, one approach is to double the sampling rate
with each iteration. Randomizing the low discrepancy sequence
freshly for each iteration, the scheme is unbiased, see Sect.~\ref{Sec:ProgressiveRand}.

A deterministic variant of the algorithm sequentially consumes the points
of the low discrepancy sequence along the space filling curve
according to the selected number of samples per pixel along the
iterations. The segment lengths drawn from the sequence then are a multiples of
the length of the selected space filling curve. In two dimensions, the length is a
quadratic power of the base of the selected space-filling curve.

For the example of base two, segment lengths amount to multiples of
powers of two. Hence, for a specific pixel, samples drawn from a radical inverse
in base $b = 2$ are spaced in multiples of powers of two, which will
reveal correlation artifacts as visualized in Fig.~\ref{Fig:Correlation}.
Depending on the combination of space filling curve and component
of the low discrepancy sequence, samples may even not be distributed
uniformly, as explained in Sect.~\ref{Sec:Correlation}.
Omitting the components of the low discrepancy sequence that correlate
with the space-filling curve may help, but as revealed in Sect.~\ref{Sec:Correlation},
there may be multiple components with correlations.

The issues of leapfrogging low discrepancy sequences have been encountered
in parallelizing quasi-Monte Carlo methods \cite{KW97,Abr02}. Their remedy
has led to the concept of partitioning low discrepancy sequences into multiple
low discrepancy sequences, which we explore next.

\subsection{Partitioning one Low Discrepancy Sequence} \label{Sec:Consistent}

Partitioning a low discrepancy sequence into a finite number of low discrepancy
sequences has been introduced in \cite{ParQMC} for the purpose of
parallel quasi-Monte Carlo methods: One dimension of a low discrepancy
sequence is used for partitioning, while the remaining dimensions are used
for quasi-Monte Carlo integration. We use the principle to develop a simple
consistent algorithm for rendering along the Hilbert curve.

Let $\phi_b(i)$ be the component of a low discrepancy sequence to be
partitioned into $N = b^m$ low discrepancy sequences and let $\vec{x}_i$
be the points of that low discrepancy sequence without the component
used for partitioning. Then the integers
\[
  \lfloor N \cdot  \phi_b(i) \rfloor
  = \lfloor b^m \cdot \phi_b(i) \rfloor
  = \left\lfloor b^m \cdot \sum_{k = 0}^\infty a_k(i) b^{- k - 1} \right\rfloor ,
\]
form a permutation of $\{0, \ldots, N - 1\}$ that repeats every
$N$ points. Selecting $N$ as the length of a space filling
curve and $b$ its base, each pixel with index $j$ along the
space-filling curve is assigned the sequence of points
\begin{equation*}
  P_j = \left\{\vec{x}_{l \cdot N + \phi_b^{-1}(j/N)}: l \in \mathbb{N}_0 \right\}
  \Leftrightarrow
  P_{\phi_b^{-1}(j/N)} = \left\{\vec{x}_{l \cdot N + j}: l \in \mathbb{N}_0 \right\}
\end{equation*}
of the original low discrepancy sequence $\vec{x}_i$, which results
in an overall consistent deterministic quasi-Monte Carlo method \cite[Sect.1.1]{NutshellQMC}.

As the offset $\phi_b^{-1}(j/N)$ is constant per pixel, omitting the inverse of the permutation
$\lfloor b^m \cdot \phi_b(i) \rfloor$ and instead assigning
\[
  P_j = \left\{\vec{x}_{l \cdot N + j}: l \in \mathbb{N}_0 \right\}
\]
simplifies the implementation: Now contiguous blocks
of the low discrepancy sequence are enumerated along the
space-filling curve and we have the additional benefit of locally
improved uniformity as described in Sect.~\ref{Sec:HilbertEnumeration}
and illustrated in Fig.~\ref{Fig:SpaceFillingCurves}.

For the combination of the Hilbert curve and the improved Halton
sequence \cite{ImprovedHalton} for progressive image synthesis,
we have $b = 2$ and use $\phi_2$ to partition the low discrepancy sequence
into $N = 2^{2n}$ low discrepancy sequences, one for each pixel
along the Hilbert curve. The algorithm uses
the pixel index $j$ along the Hilbert curve as offset into the
low discrepancy sequence $\vec{x}_i$ and leapfrogs from
there with a stride of $N$, which is simple to execute on a
massively parallel computer system, like for example a GPU cluster
\cite{ParQMC,Iray_report,RTG:LoadBalancing}.

\section{Results and Discussion} \label{Sec:Discussion}

The Hilbert curve has been applied in computer graphics before.
Rendering images by enumerating pixels along the Hilbert curve
improves performance by higher cache hit rates due to the locality
properties of the Hilbert curve. The visual benefits
of half-toning by dithering along the Hilbert curve have been recognized
in \cite{HilbertHalftoning}.

Our new algorithms benefit from
these findings. They are a special case of Array-RQMC
that does not require sorting, as we rely on the bijection between pixels
and the space-filling curves. As indices can be computed directly,
neither lookup tables nor additional memory for lookup tables are required.

\begin{figure}
  \centering
  {
    \def\arraystretch{2}\tabcolsep=0pt
    \begin{tabular}{@{}ccc@{}}
      \multicolumn{3}{c}{%
      \begin{tikzpicture}
        \node[anchor = north west, inner sep = 0, outer sep = 0] at (0, 0) {%
          \includegraphics[width=\textwidth]{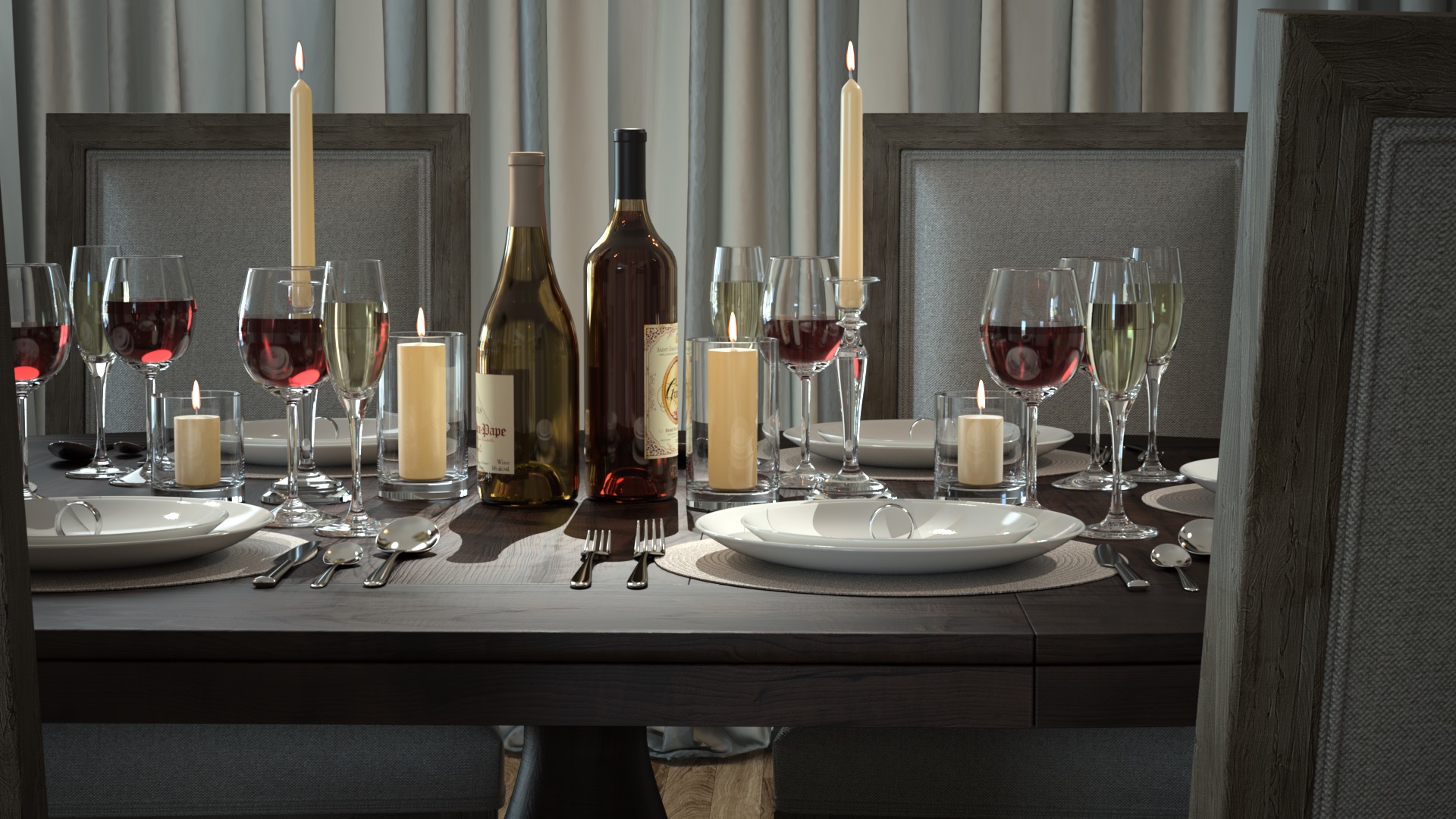}%
        };
        \draw[red, very thick] (8*0.609, -7.5*0.609) rectangle ++(4*0.609, -1.2*0.609);
      \end{tikzpicture}
      }\\
      \includegraphics[width=.32\textwidth]{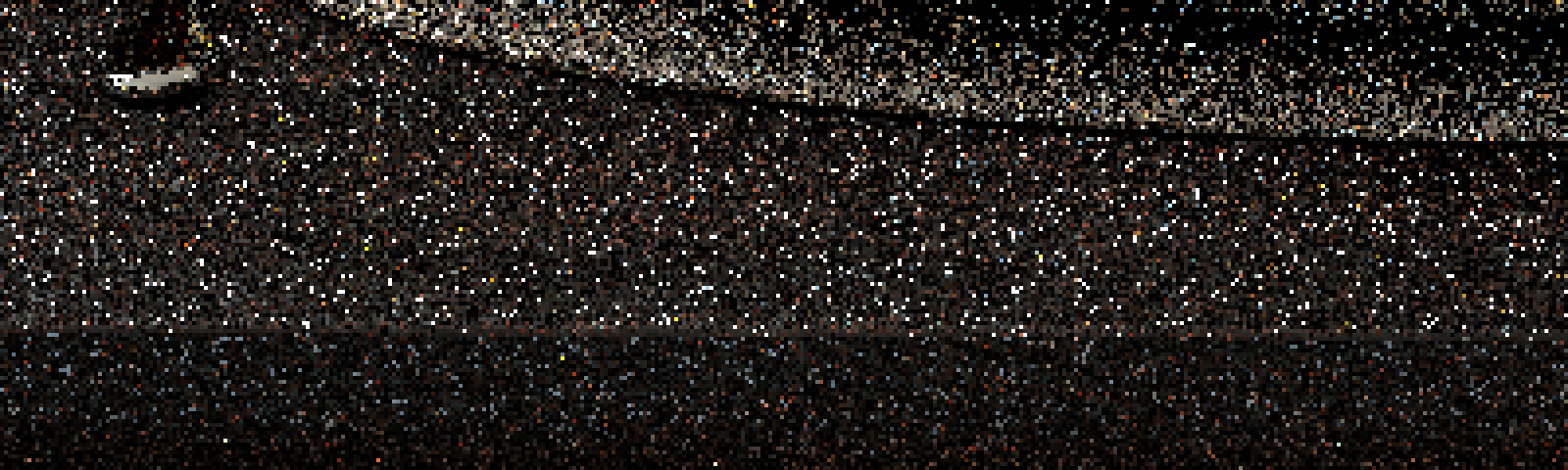}\hspace*{.02\textwidth}&
      \includegraphics[width=.32\textwidth]{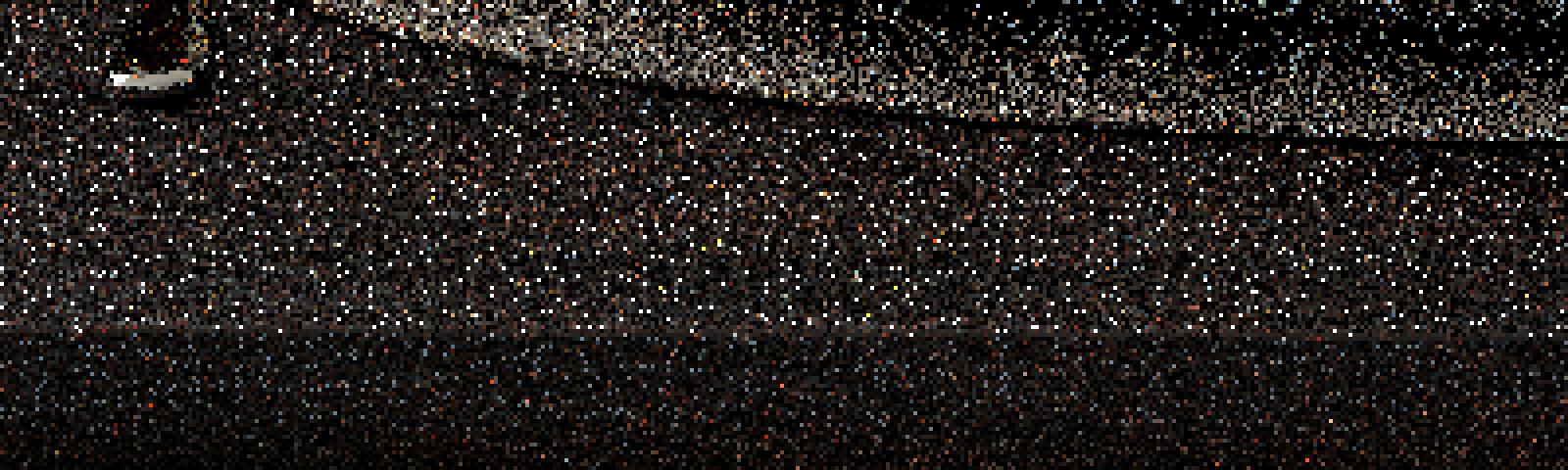}\hspace*{.02\textwidth}&
      \includegraphics[width=.32\textwidth]{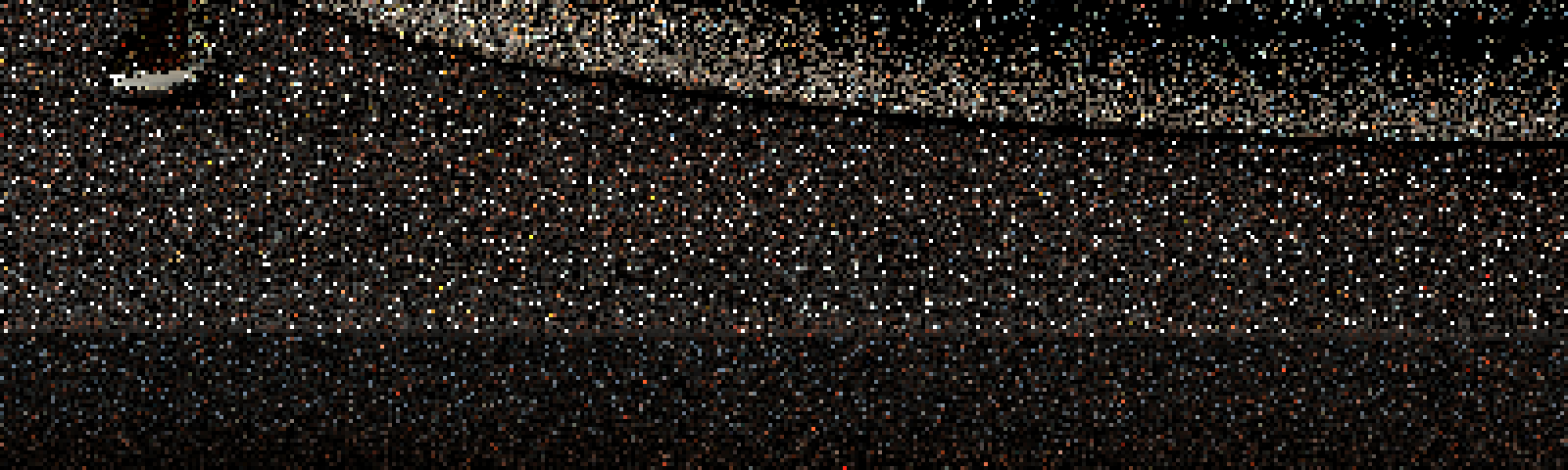}\\
      \includegraphics[width=.32\textwidth]{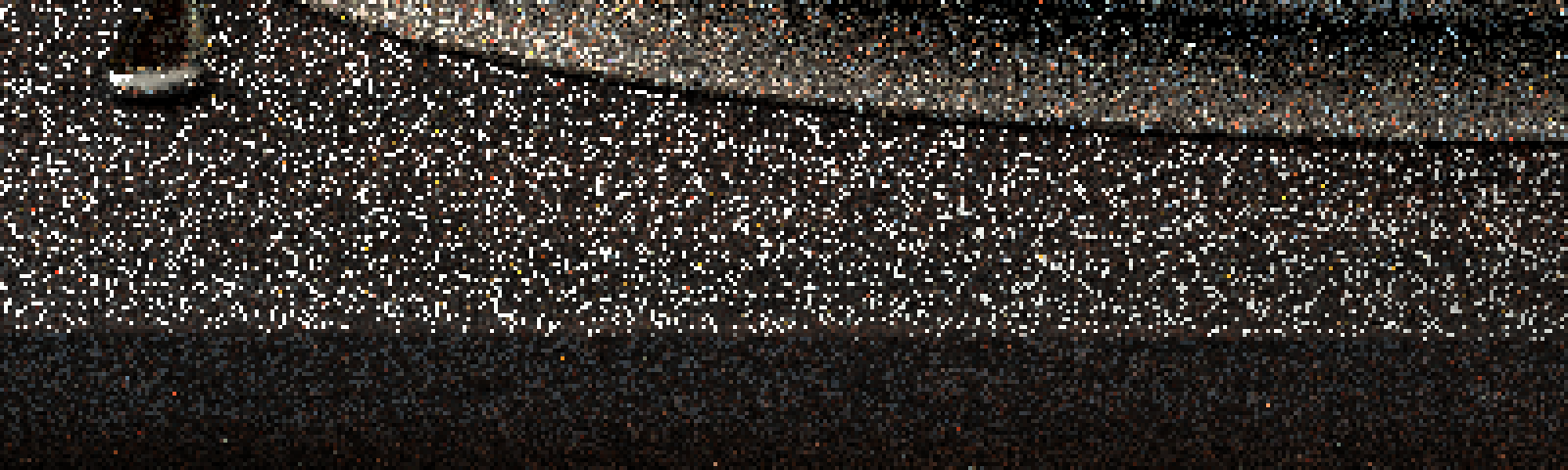}\hspace*{.02\textwidth}&
      \includegraphics[width=.32\textwidth]{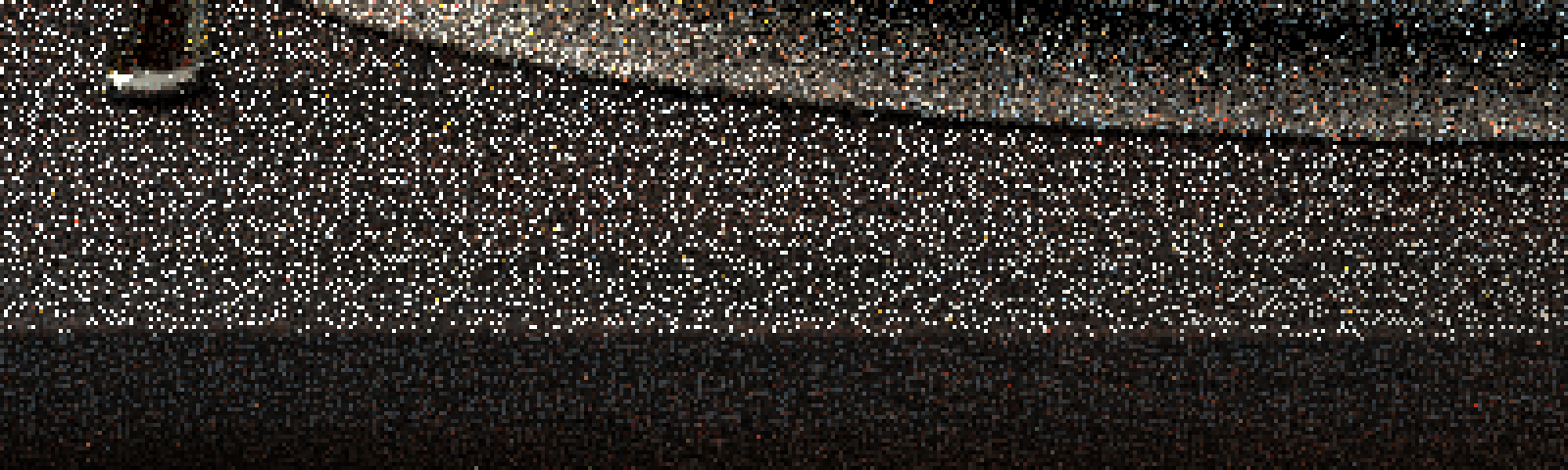}\hspace*{.02\textwidth}&
      \includegraphics[width=.32\textwidth]{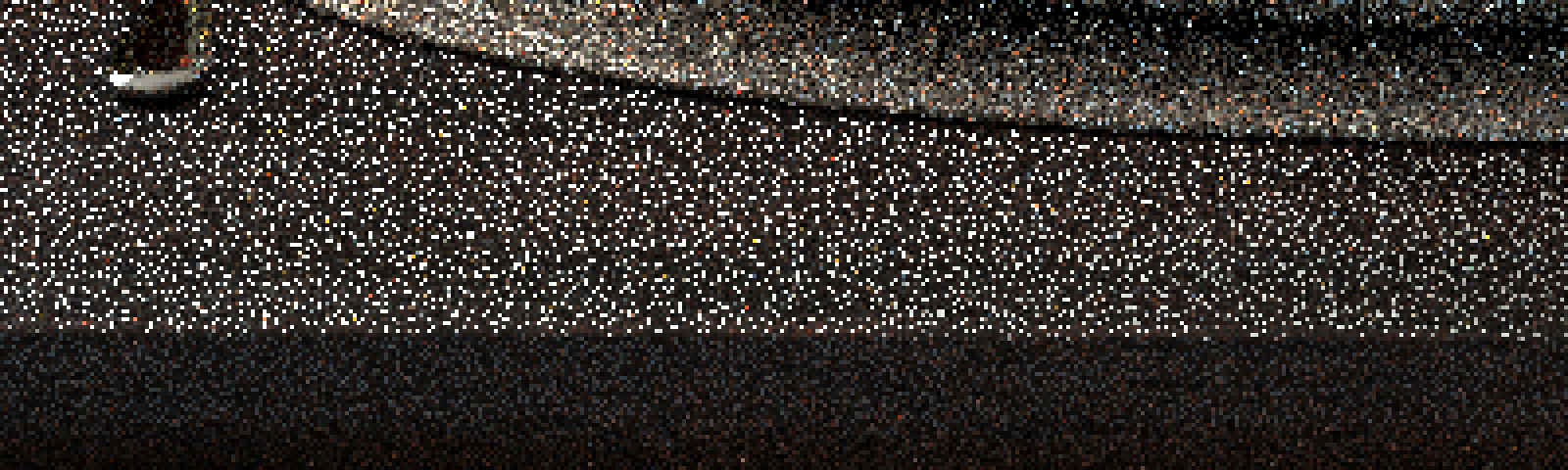}\\
      \includegraphics[width=.32\textwidth]{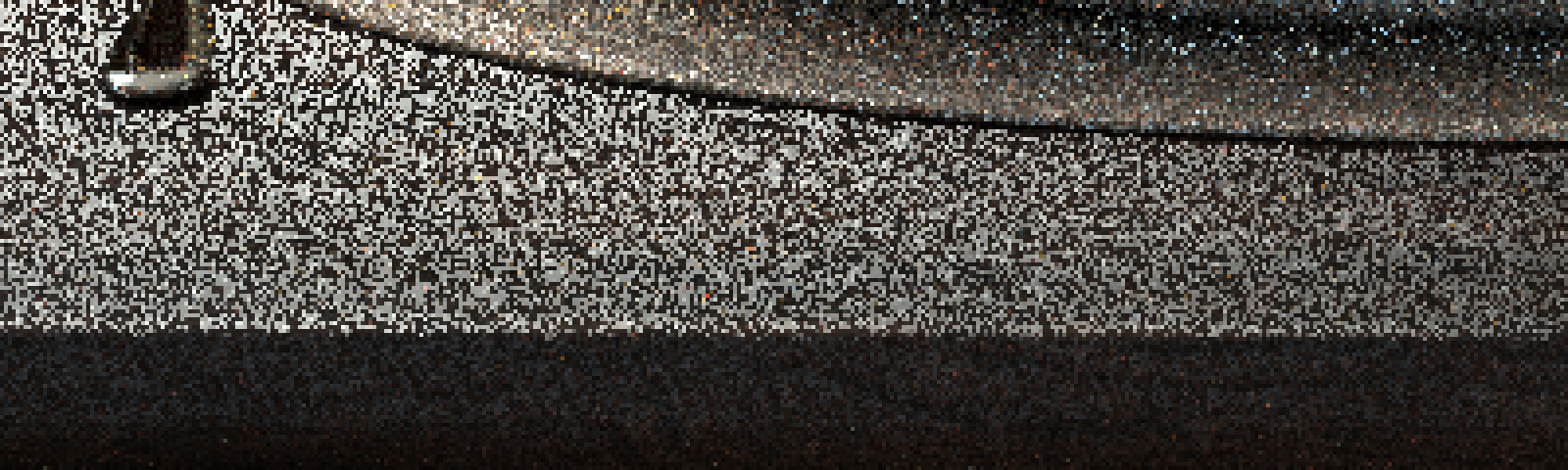}\hspace*{.02\textwidth}&
      \includegraphics[width=.32\textwidth]{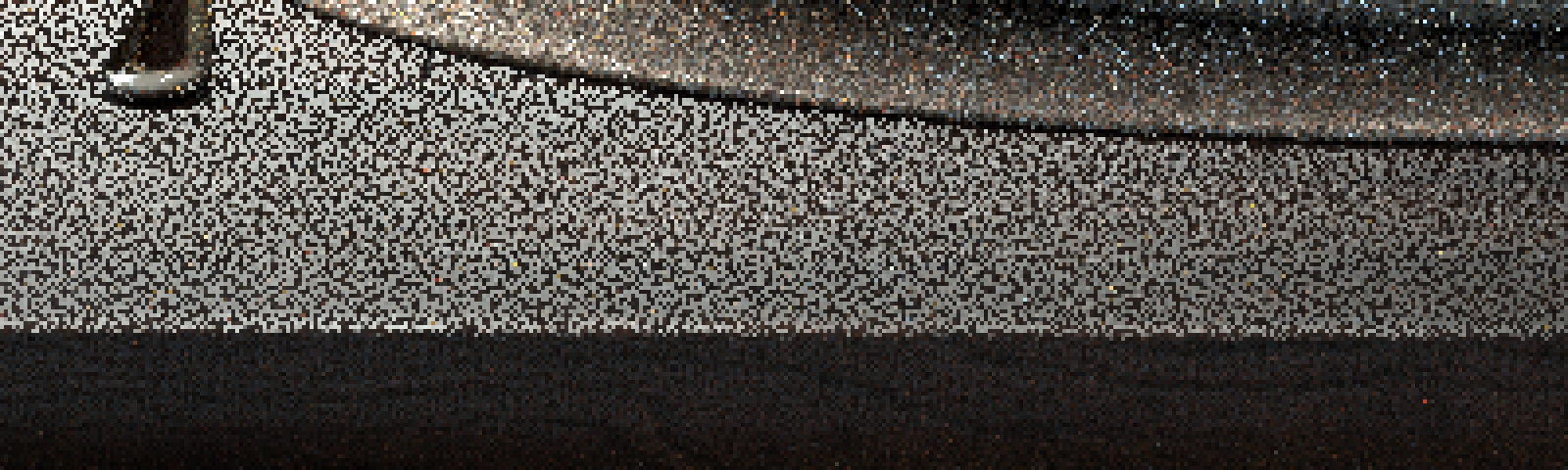}\hspace*{.02\textwidth}&
      \includegraphics[width=.32\textwidth]{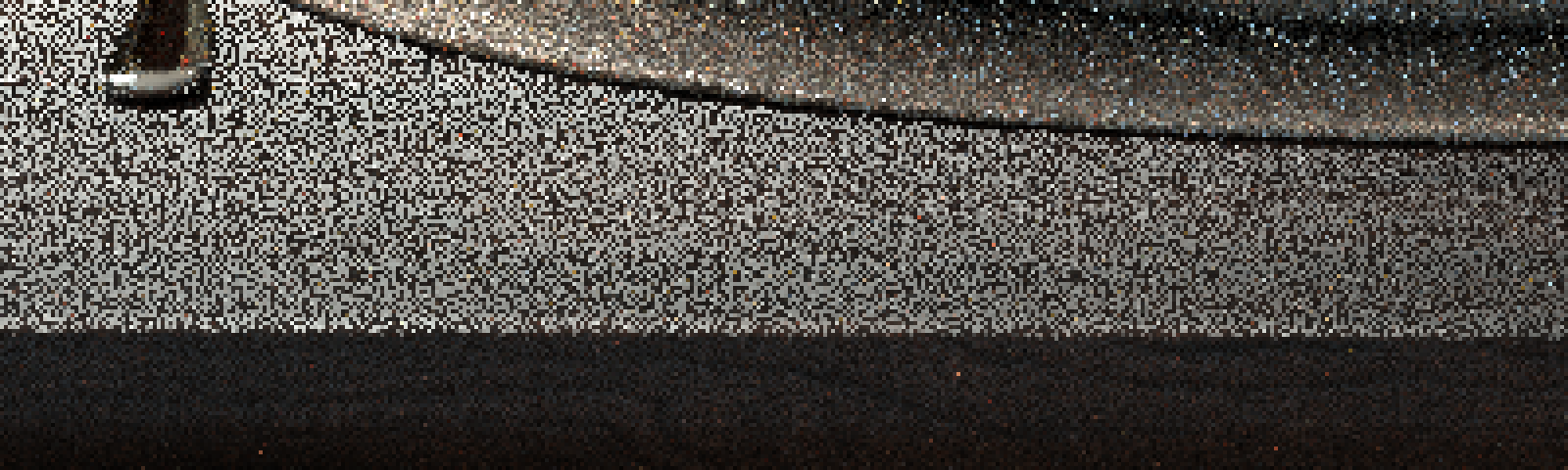}\\
      \includegraphics[width=.32\textwidth]{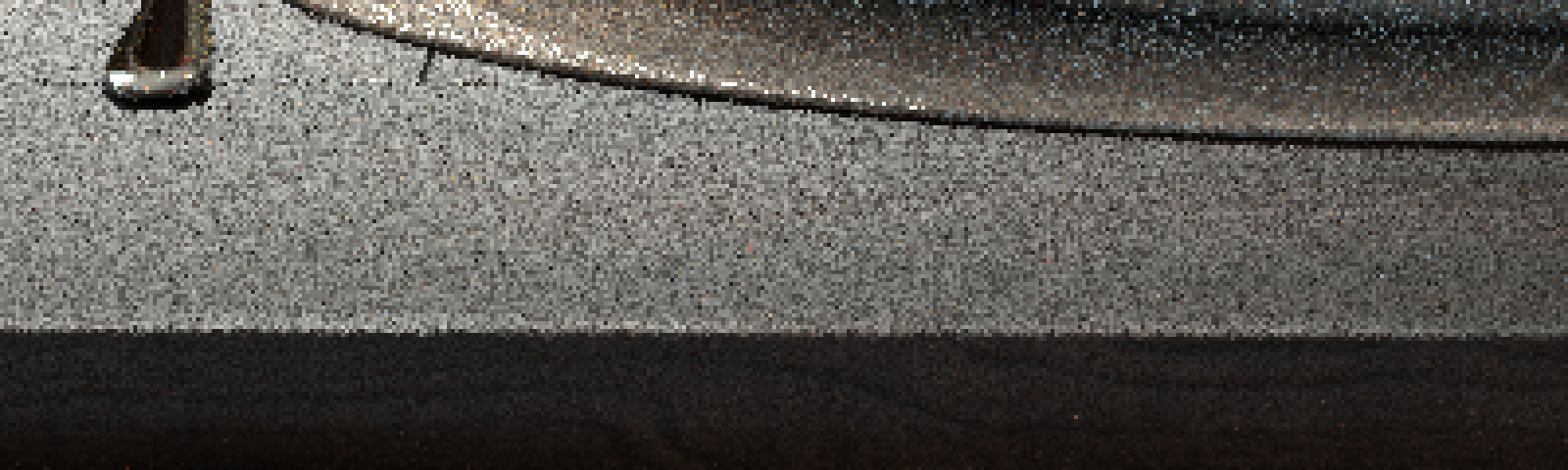}\hspace*{.02\textwidth}&
      \includegraphics[width=.32\textwidth]{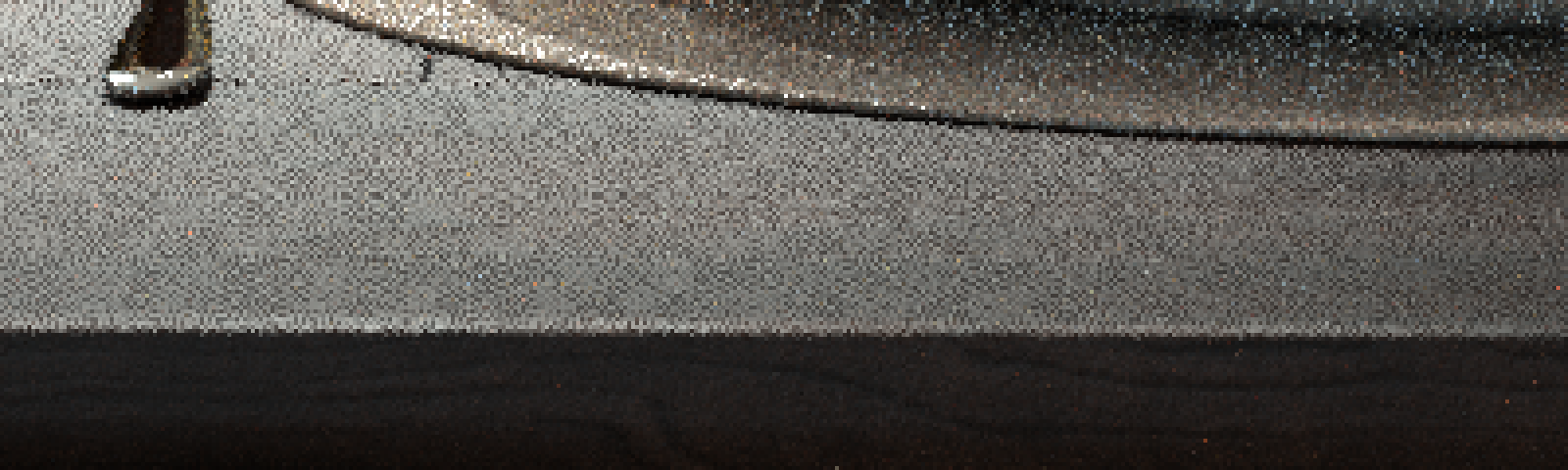}\hspace*{.02\textwidth}&
      \includegraphics[width=.32\textwidth]{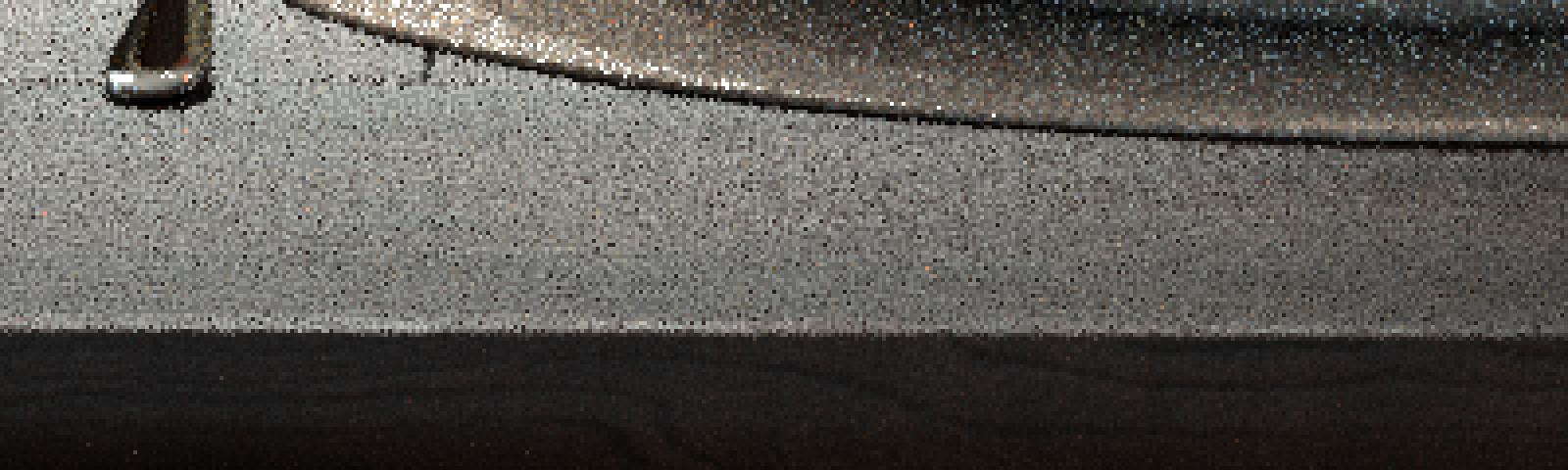}\\
      \shortstack{Halton sequence\\across image plane}&
      \shortstack{Extensible lattice shifted by\\ Halton along Hilbert curve}&
      \shortstack{Halton sequence\\along Hilbert curve}
    \end{tabular}
  }
  \caption{Photorealistic image synthesis using one low discrepancy sequence
  across the image plane (left column, \cite{SampleEnum}),
  drawing samples from one low discrepancy sequence while enumerating the pixels along the Hilbert curve
  to shift an extensible lattice for sampling inside a pixel (middle column, Sect.~\ref{Sec:DeterministicShift})
  and in contiguous blocks to directly sample inside a pixels (right column, Sect.~\ref{Sec:Consistent}).
  The top image has been rendered using 100,000 samples per pixel, while the
  insets from top to bottom were rendered at 1, 4, 16, and 64 samples per pixel, respectively.
  The more uniformly distributed and less splotchy appearance of the sampling noise
  is especially visible in areas that at higher sampling rates appear smooth, like the table top or the back rests.
  The difference in quality is clearly visible on a computer screen and may be difficult
  to reproduce in print. The reader may need to vary the distance of observation.
  As all methods are consistent, the observable differences vanish with an increasing number of samples per pixel.
  Nevertheless, the improvement very much matters in settings, where only a small number of samples are
  affordable, such as in real-time rendering.
}
  \label{Fig:FiniteSamples}
\end{figure}

Our focus is on deterministic algorithms, as these can be reliably
parallelized and results are exactly reproducible \cite{NutshellQMC}.
We use an improved variant of the Halton sequence
\cite{ImprovedHalton} in the experiments.
The implementation of fitting elementary intervals to the pixel raster \cite{SampleEnum}
is involved and requires 64-bit signed integers to run Euclid's algorithm
for the Chinese remainder theorem. This computation is not required for
the algorithms in Sect.~\ref{Sec:HilbertEnumeration} and Sect.~\ref{Sec:Consistent}
that are straightforward to implement. Shifting a rank-1 lattice sequence \cite{eLattices:01} by
the Halton sequence enumerated along the Hilbert curve (see Sect.~\ref{Sec:DeterministicShift})
is even simpler and practical with only two 32-bit integer indices. We employ an extensible lattice
constructed by primitive polynomials \cite{PolyR1LS}.
To assess the visual differences of the classic \cite{SampleEnum} and the two new progressive
sampling approaches, their results are compared at
low sampling rates in Fig.~\ref{Fig:FiniteSamples} .

The most prominent advantage of the new algorithms
is inherited from Array-RQMC: As illustrated in Fig.~\ref{Fig:SpaceFillingCurves},
the Hilbert curve has the smallest number of curve segments in the neighborhood
around a pixel. Hence more consecutive samples are used locally, which makes
better use of the uniformity of a low discrepancy sequence across pixels
as compared to other space-filling curves. Hence, the noise in the images is more
uniformly distributed noise at low sampling rates.

Our methods achieve a quality comparable to methods that require optimization
\cite{DitheredSampling,ScreenSpaceBlueNoise}, are available
for any number of dimensions, are simpler than other approaches
that sample along space-filling curves \cite{Z-Sampler}, are
deterministic, and are consistent.

We have not yet explored the potential of selecting or reordering the dimensions of low discrepancy sequences. This is
an interesting direction of future research that has been initially explored
for rank-1 lattices in computer graphics \cite{R1Lpathintegral}.
Furthermore, it is worth investigating the many other possible combinations
of low discrepancy sequences and space-filling curves with respect
to their visual quality and convergence speed.

\section{Conclusion}

Based on the seminal work on Array-RQMC \cite{SortingArrayQMC}, we introduced
simple deterministic consistent rendering algorithms that at low sampling rates
produce noise characteristics that are very amenable to the human eye.
Key to the algorithms are the preservation of discrepancy when
enumerating low discrepancy sequences along the Hilbert curve and
the principle of partitioning one low discrepancy sequence into multiple.
It appears that the correlation of samples across pixels
via low discrepancy may be more relevant to the eye than their independence.

\end{document}